%% file: main.tex
\documentclass{article}
\usepackage{microtype}
\usepackage{graphicx}
\usepackage{booktabs}
\usepackage{amsmath,amssymb,amsfonts}
\usepackage{amsthm}
\usepackage{tabularx}
\usepackage{array}
\usepackage{xcolor}
\usepackage{url}
\usepackage{hyperref}
\hypersetup{pdfauthor={Tianyu Ding, Yaocheng Tong, Muzhi Yang, Lei Yang, Fei Jiang},pdftitle={Testing Processor Contracts in Speculative Decoding Engines},pdfsubject={Research preprint}}
\makeatletter
\g@addto@macro\UrlBreaks{\do\a\do\b\do\c\do\d\do\e\do\f\do\g\do\h\do\i\do\j\do\k\do\l\do\m\do\n\do\o\do\p\do\q\do\r\do\s\do\t\do\u\do\v\do\w\do\x\do\y\do\z\do\0\do\1\do\2\do\3\do\4\do\5\do\6\do\7\do\8\do\9}
\makeatother

\usepackage[accepted]{mlsys2025}
\makeatletter
\renewcommand{\Notice@String}{Preprint.}
\makeatother
\hypersetup{pdfauthor={Tianyu Ding, Yaocheng Tong, Muzhi Yang, Lei Yang, Fei Jiang},pdfsubject={Research preprint}}
\mlsystitlerunning{Testing Processor Contracts in Speculative Decoding Engines}
\input{tables/numbers.tex}
\input{tables/numbers_m1.tex}
\input{tables/numbers_m2.tex}
\input{tables/numbers_m3.tex}
\input{tables/numbers_m4.tex}

\newcommand{\badc}{\textsc{Badc}}
\newcommand{\Law}{\mathrm{Law}}
\newtheorem{lemma}{Lemma}
\newtheorem{lemma-plain}{Lemma}
\begin{document}
\twocolumn[
\mlsystitle{Testing Processor Contracts in Speculative Decoding Engines}
\begin{mlsysauthorlist}
\mlsysauthor{Tianyu Ding}{ind}
\mlsysauthor{Yaocheng Tong}{ind}
\mlsysauthor{Muzhi Yang}{ind}
\mlsysauthor{Lei Yang}{ind}
\mlsysauthor{Fei Jiang}{ind}
\end{mlsysauthorlist}
\mlsysaffiliation{ind}{Independent Researcher}
\mlsyscorrespondingauthor{Tianyu Ding}{tid16@pitt.edu}
\mlsyskeywords{speculative decoding, conformance testing, test oracles, statistical testing, language-model inference engines}
\vskip 0.3in
\begin{abstract}
\input{sections/00-abstract}
\end{abstract}
]
\printAffiliationsAndNotice{}
\input{sections/01-intro}
\input{sections/02-background}
\input{sections/03-problem}
\input{sections/04-approach}
\input{sections/05-design}
\input{sections/06-results}

\input{sections/07-discussion}
\input{sections/08-related}
\input{sections/09-conclusion}
\input{sections/10-data}
\clearpage
\bibliographystyle{mlsys2025}
\bibliography{references}
\clearpage
\input{sections/11-ai-usage}
\paragraph*{Author statement}
The authors contributed to this work in their personal capacities as independent researchers.
The views expressed are their own and do not necessarily reflect those of their current or former employers.

\clearpage
\appendix
\makeatletter
% Layout fix: when \clearpage sends a pending two-column table to a float page, the LaTeX kernel
% keeps \@colroom at the reduced height, so the next column comes out short. Reset it.
\g@addto@macro\@doclearpage{\global\@colroom\@colht}
\makeatother
\section*{Supplementary Material}
The appendices retain the supplementary proof, study details and limitations.
Supplementary tables, figures and equations are numbered separately from the main text.
\setcounter{table}{0}
\setcounter{figure}{0}
\setcounter{equation}{0}
\renewcommand{\thetable}{S\arabic{table}}
\renewcommand{\thefigure}{S\arabic{figure}}
\renewcommand{\theequation}{S\arabic{equation}}
\providecommand{\theHtable}{}
\providecommand{\theHfigure}{}
\providecommand{\theHequation}{}
\renewcommand{\theHtable}{supp.\arabic{table}}
\renewcommand{\theHfigure}{supp.\arabic{figure}}
\renewcommand{\theHequation}{supp.\arabic{equation}}
\input{supplement/arxiv-restatement}

\input{supplement/proof-body-mlsys}
\input{supplement/additional-details}

\end{document}

%% file: tables/numbers.tex
\newcommand{\TvThreshold}{0.01}
\newcommand{\CalDelta}{0.005}
\newcommand{\RepairGap}{0.01}
\newcommand{\RepairRate}{10^{-4}}
\newcommand{\NumConfigs}{26}
\newcommand{\NumControls}{7}
\newcommand{\NumProbes}{19}
\newcommand{\Mstage}{26}
\newcommand{\NperArm}{200{,}000}
\newcommand{\NumBuilds}{5}
\newcommand{\VocabHF}{7}
\newcommand{\LHF}{3}
\newcommand{\VocabLCP}{10}
\newcommand{\LLCP}{4}
\newcommand{\DraftHF}{3}
\newcommand{\NgramHF}{2}
\newcommand{\DraftLCP}{4}
\newcommand{\TotalDraws}{17.8}
\newcommand{\CoreHours}{33.7}
\newcommand{\PropHFOneSevenNZero}{927{,}150}
\newcommand{\DHFOneSevenSix}{0.0142}
\newcommand{\TvHHFOneSevenSix}{0.022}
\newcommand{\DHFOneSevenSeven}{0.6417}
\newcommand{\TvHHFOneSevenSeven}{0.647}
\newcommand{\DHFXEight}{0.4060}
\newcommand{\DnHFXEight}{0.0091}
\newcommand{\ErrHFXEight}{48{,}543}
\newcommand{\WitLCPOneOneNine}{23{,}335}
\newcommand{\DLCPOneOneNine}{0.1131}
\newcommand{\TvHLCPOneOneNine}{0.117}
\newcommand{\WitLCPOneThreeFive}{23{,}407}
\newcommand{\DLCPOneThreeFive}{0.1135}
\newcommand{\DLCPTwoOneNine}{0.1490}
\newcommand{\TvHLCPTwoOneNine}{0.156}
\newcommand{\DLCPTwoThreeFive}{0.1475}
\newcommand{\DLCPThreeOneNine}{0.0078}
\newcommand{\TvHLCPThreeOneNine}{0.009}
\newcommand{\AccHFNZero}{29{,}054}
\newcommand{\MinTokensM}{3}
\newcommand{\NumCausesW}{two}
\newcommand{\NumCausesWCap}{Two}
\newcommand{\NumNegControlsW}{five}
\newcommand{\NumPosControlsW}{two}
\newcommand{\NumEligibleFailW}{six}
\newcommand{\NumEligibleFailWCap}{Six}
\newcommand{\NumConfirmW}{three}
\newcommand{\TrlCap}{3}
\newcommand{\RecHFOneSevenSix}{0.0195}
\newcommand{\GapHFOneSevenSix}{0.0032}
\newcommand{\NRepHFOneSevenSix}{1{,}000{,}000}
\newcommand{\RecLCPTwoOneNine}{0.1549}
\newcommand{\GapLCPTwoOneNine}{0.0020}
\newcommand{\NSThreeLCPTwoOneNine}{1{,}000{,}000}
\newcommand{\NRepLCPTwoOneNine}{200{,}000}
\newcommand{\RecLCPTwoThreeFive}{0.1548}
\newcommand{\GapLCPTwoThreeFive}{0.0023}
\newcommand{\SThreeM}{3}
\newcommand{\MdeMin}{0.011}
\newcommand{\MdeMax}{0.022}
\newcommand{\DnoAHHFOneSevenSix}{0.0017}
\newcommand{\NumEligNoAHW}{five}
\newcommand{\QueriesA}{32}
\newcommand{\QueriesB}{46}

%% file: tables/numbers_m1.tex
\newcommand{\MOneBLCPNineteenDet}{6}
\newcommand{\MOneBLCPThirtyFiveDet}{6}
\newcommand{\MOneBHFGuidDet}{4}
\newcommand{\MOneBHFGuidAtt}{3}
\newcommand{\MOneBK}{6}
\newcommand{\MOneCPass}{14}

%% file: tables/numbers_m3.tex
\newcommand{\MThreeCampaigns}{5}

\newcommand{\MThreeExposed}{52}
\newcommand{\MThreeExposedD}{29}
\newcommand{\MThreeExposedU}{23}

\newcommand{\MThreeFallbackN}{20}
\newcommand{\MThreeGPUh}{8.3}
\newcommand{\MThreeMissingDisc}{2}
\newcommand{\MThreePlanned}{60}

\newcommand{\MThreeRedet}{28}

\newcommand{\MThreeRedetReq}{29}
\newcommand{\MThreeUnexposed}{6}
\newcommand{\MThreeUnexposedDet}{3}

\newcommand{\MThreeVZeroCtlDet}{4}
\newcommand{\MThreeVZeroCtlRej}{5}
\newcommand{\MThreeVZeroTtpRej}{4}

%% file: tables/numbers_m4.tex
\newcommand{\MFourCampaigns}{10}

\newcommand{\MFourCoreK}{868}

\newcommand{\MFourdACP}{0.26--0.39}

\newcommand{\MFourdAZeros}{9--10}
\newcommand{\MFourdBDZeros}{9--10}
\newcommand{\MFourdBUZeros}{9}

%% file: sections/00-abstract.tex
Speculative sampling can accelerate decoding without changing the target distribution. Its standard guarantee
assumes that logits processors, which enforce constraints or modify token scores, receive the appropriate
token prefix and state at each verification position. We formalize this processor-prefix contract and present
\badc{} (branch-aware differential conformance testing). \badc{} combines enumerated references on tiny random
models, checks of processor inputs and state, and simultaneous confidence bounds on output-event probabilities.
Preregistered, source-informed experiments on selected Transformers and llama-cpp-python releases confirm
\NumCausesW{} causes: processors read unverified drafts or advance state without rollback. For one pilot-selected prompt on a trained checkpoint, fresh samples give a 60.26-percentage-point lower bound on the difference
in the probability of a predefined first-token event between ordinary and prompt-lookup decoding, with
nominal simultaneous 95\% coverage under within-arm i.i.d.\ sampling. Controlled comparisons establish
no cause-recovery advantage of directed selection over random selection, while permutation testing detects each
tested family at least as often as \badc{}. These findings connect processor-interface failures to observable
decoding differences without establishing diagnostic superiority or workload prevalence.

%% file: sections/01-intro.tex
\section{Introduction}
\label{sec:intro}

Speculative sampling preserves the target distribution under algorithmic assumptions~\cite{leviathan2023,chen2023specsampling},
including use of the processed target distribution at each verification position's logical prefix
(Section~\ref{sec:problem}). Logits processors modify token scores to enforce constraints or apply guidance.
Even when the model logits are correct, a speculative implementation can give these processors the wrong inputs:
a processor may read unverified draft tokens, or retain state from calls that sequential decoding would not make.
We test this processor-prefix contract, together with stopping rules and probability records, in fixed engine builds.

\input{figures/fig-overview-float}

Equal token support, frequency ordering and greedy outputs do not establish equal sampling
distributions~\cite{batchspec2025,greedyspec2026,greedycorrect2026}.
Reinforcement-learning (RL) rollout systems rely on distribution preservation~\cite{nemorl2026,bubblespec2026,efficientrollout2026,adaptivedraft2025,deepseekv41flash2026}.
We test the engine boundary without a trainer--engine run or measured RL consequences.

\badc{} (branch-aware differential conformance testing) compares returned-output distributions with speculation
enabled and disabled for deterministic drafters, including prompt lookup (Fig.~\ref{fig:overview}).
On tiny random models, enumeration supplies a sequential reference for calibration, processor-contract checks,
probability records and source-derived mechanism hypotheses. Fresh repaired-engine samples then bound remaining
differences on events fixed before confirmation. The reference depends on adapters and floating-point evaluation;
compatibility with a mechanism hypothesis does not uniquely identify the mechanism.

The preregistered study tests \NumConfigs{} configurations on \NumBuilds{} fixed
Transformers~\cite{wolf2020transformers} and llama-cpp-python~\cite{sw-llamacpppython} builds
(Sections~\ref{sec:protocol},~\ref{sec:design}).
RQ1--RQ4 cover calibration, mechanisms, controlled comparisons and transfer to historical
vLLM defects~\cite{kwon2023vllm,sw-vllm}.

Our contributions are:
\begin{itemize}
\item A processor-prefix contract and conditional composition result, together with executable mechanism
tests and simultaneous bounds for predefined output events (Sections~\ref{sec:problem}--\ref{sec:approach}).
\item Evidence that draft exposure and retained processor state can change sampling distributions, with
\NumCausesW{} source-linked, intervention-confirmed causes. Held-out tiny-model tests and bounded
trained-checkpoint guidance studies examine transfer (Sections~\ref{sec:rq1}--\ref{sec:rq2}).
\item Controlled comparisons showing no cause-recovery advantage from directed selection, plus unchanged-rule
detection on public historical vLLM families (Sections~\ref{sec:rq3}--\ref{sec:rq4}).
\end{itemize}

%% file: figures/fig-overview-float.tex
\begin{figure*}[t]
\centering
% generated by paper-icst/scripts/make_diagrams.py (ARIS FigureSpec: figures/specs/fig-overview.json, source
% figures/src/fig-overview.svg); 1350 x 250 px canvas at \textwidth: text prints at 7.26 pt and 7.64 pt
\includegraphics[width=\textwidth]{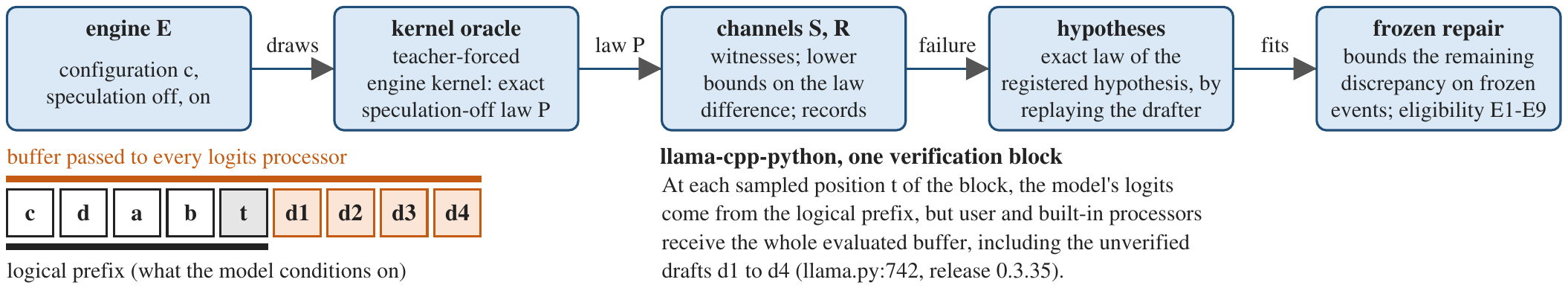}
\caption{\badc{} pipeline (top) and the llama-cpp-python mechanism confirmed in Section~\ref{sec:rq2} (bottom).
Both arms run the unmodified engine; for vLLM the oracle is a CPU reference (Section~\ref{sec:rq4}).}
\label{fig:overview}
\end{figure*}

%% file: sections/02-background.tex
\section{Background and Motivating Example}
\label{sec:background}

\paragraph{Speculative sampling}
\looseness=-1
A drafter proposes $k$ tokens $d_1,\dots,d_k$ with proposal probabilities $q_i$, and the target
model scores them in one forward pass. Token $d_i$ is accepted with probability
$\min(1, p_i(d_i)/q_i(d_i))$, where $p_i$ is the target distribution at that position; at the
first rejection the engine samples from the normalized residual $\max(p_i-q_i,0)$, and a fully
accepted block adds
a bonus token from $p_{k+1}$. The output is distributed exactly as sequential sampling from
$p$~\cite{leviathan2023,chen2023specsampling}. This is equality of output distributions, not a requirement that the
drafter match the target model or that a shared seed produce identical outputs. \emph{Model-free} drafters copy text instead of
running a draft model~\cite{llma2023}; the prompt-lookup drafters we test copy the continuation of
an $n$-gram match in the prompt and generated text. Their proposal is a point mass, so $d_i$ is accepted
with probability $p_i(d_i)$, which Transformers and llama-cpp-python implement by
\emph{sample-and-match}: sample $y_i\sim p_i$ and accept while $y_i=d_i$.

\paragraph{What engines compose with the sampler}
The target distribution in this guarantee is the \emph{processed} distribution, after temperature,
truncation, penalties, biases, minimum length, grammar masks, guidance~\cite{sanchez2023cfg} or
watermarking~\cite{dathathri2024synthid}. Engines also stop generation and record probabilities. During speculation, these components may
receive uncommitted tokens, including drafts later rejected.

\paragraph{Why RL rollouts care}
PPO and GRPO use rollout samples and policy probabilities~\cite{schulman2017ppo,shao2024grpo};
TRL corrects engine--trainer mismatch with recorded-log-probability ratios~\cite{sw-trl}.

\paragraph{Motivating example}
llama-cpp-python's prompt-lookup loop~\cite{sw-llamacpppython} scores positions at their logical prefixes,
but its processor callback reads \texttt{self.\_input\_ids} (\texttt{llama\_cpp/llama.py}, line~742 in 0.3.35),
including unverified drafts (Fig.~\ref{fig:overview}). The built-in \texttt{MinTokensLogitsProcessor} counts
these drafts as generated tokens, releasing the end-of-sequence mask early. After the first callback fixes
prompt length, a block one committed token short of the minimum should mask end-of-sequence at its first
position; drafts make the buffer appear long enough. The frozen repair passes the logical prefix
(Section~\ref{sec:rq2}). Short visible completions alone are not witnesses, because the API hides internal tokens.

%% file: sections/03-problem.tex
\section{Problem Formulation}
\label{sec:problem}

Fix an engine $E$ at a pinned version, a model, a prompt $x$, a horizon $L$, and a sampling
configuration $c$.

\paragraph{Kernel, trace, and return map}
The \emph{sequential kernel} $K_E(\cdot \mid h)$ is the processed next-token distribution that $E$
samples from in sequential decoding after the committed prefix $h$, with every stateful processor replayed along $h$ from its initial state.
Sequential decoding yields a trace $T=(t_1,\dots,t_n,\omega)$ with termination code $\omega$; the
engine returns $\rho_E(T)$, where the \emph{return map} $\rho_E$ keeps what the caller sees (token
identifiers and any finish reason). The speculation-off law is the push-forward
$P=\rho_E\#\Law_K(T)$, and $\mu$ is the speculation-on law, both of returned outputs. The record $\ell_t$ is the log-probability reported or implied for token $y_t$, with documented value $r_E(h,y_t)$ at the logical prefix $h$. We do not claim that $P$ equals a consumer's policy
$\pi_\theta$.

\paragraph{When speculation is lossless}
Four sampler conditions make speculative sampling exact for the model-free drafters we test:
(i)~acceptance uses the proposal law actually used; (ii)~acceptance, residual and bonus steps are
implemented correctly; (iii)~the verification randomness of each block position is fresh:
independent across positions and of logically earlier positions' randomness, and nothing decided
before the position (its target, its draft, whether it is reached) depends on it, though the
engine may draw it early; (iv)~the draft is a deterministic function of the committed prefix, so
the proposal is a point mass. Throughout, processors are deterministic in their inputs and state,
and the stop rule is deterministic in the committed prefix (callbacks that consult external state
are excluded). A position's
\emph{logical prefix} is the prompt, the tokens committed before its block, and the block's drafts
before it; the \emph{evaluation buffer} is every token fed to the model for the block, unverified
drafts included.

\begin{lemma}[Conditional composition]\label{lem:composition}
Assume (i)--(iv) and the following. Every reached verification position, including the bonus
position, uses the target distribution $K_E(\cdot\mid h)$ at its logical prefix $h$; for a
stateful processor, a state that is observationally indistinguishable from the sequential state at $h$
suffices. Before every emission, including the first, the sequential stop rule is applied to the
committed prefix: the engine emits the next token if and only if the rule continues, otherwise it stops and
records the rule's termination code; tokens evaluated beyond a stop are discarded. The returned output equals
$\rho_E$ of the committed trace. Then $\mu=P$.
\end{lemma}

\noindent\emph{Proof sketch.} Induct over committed positions: the emitting position's logical
prefix is the committed prefix, its target and draft are fixed by earlier positions' randomness,
and its own randomness is fresh, so the per-position speculative-sampling theorem gives the sequential conditional;
$\rho_E$ is deterministic in the committed trace. We claim no converse; the counterexample and full proof are in
Appendices A and B, respectively.

\paragraph{Defect classes}
The lemma's prefix condition is that theorem's premise. For processors it holds under the \emph{processor-prefix contract}: at every verified position the
processor sees that position's logical prefix, in a state indistinguishable from the sequential one. It suggests
three mechanism families by which $\mu$ can depart from $P$ (D1--D3), used as executable hypotheses, not as
necessary conditions; D4 covers records. \textbf{D1 (context):} a processor, penalty, or constraint reads too long a prefix (the evaluation buffer) or too short a one (a committed prefix that misses the position's own drafts).
\textbf{D2 (state):} a stateful processor advances on calls that sequential decoding never makes (rejected drafts,
or the drafter's calls on placeholder logits) without rollback, so its state reflects positions never committed. \textbf{D3 (stopping):} a
stop rule is evaluated per block, so tokens after a stop are returned or a limit is overshot.
\textbf{D4 (records):} the returned record differs from $r_E$; it changes the records, not $\mu$.

\paragraph{Test goal}
For each configuration, compare the speculation-on and speculation-off returned-output laws using $N$ draws per arm,
and check records against their documented values. The enumerated reference supplies witnesses, events, mechanism predictions and calibration.
\badc{} also tests supplied mechanisms and bounds a frozen repair's remaining discrepancy on registered events (Section~\ref{sec:approach}). Failure to reject is reported as \emph{not rejected}, never as conformance.

%% file: sections/04-approach.tex
\section{\badc{}: Branch-Aware Differential Conformance Testing}
\label{sec:approach}

\subsection{Kernel Oracle and Adapter Contract}
\label{sec:oracle}

The oracle enumerates the distribution of sequential sampling on tiny random models
(vocabularies of \VocabHF{} to \VocabLCP{} tokens, horizon \LHF{} or \LLCP{}). An engine-specific adapter matches
the sampling configuration, replays stateful processors at each logical prefix and reproduces the
stopping behavior. The supplement lists the adapter conditions (A1)--(A4).

The Transformers adapter reuses the processor and
stopping objects that \texttt{generate()} prepares; the llama-cpp-python (LCP) adapter mirrors the vendored C sampler
chain for the probed features, with every mirrored line recorded. Enumeration yields $P$ and the kernel
log-probabilities $\log K_E(y_t\mid h)$.

Exactness is relative to these mirrors and floating-point evaluation. A configuration is void if
(A5)~\emph{calibration} finds a speculation-off contract violation or an event with
$L_{\mathrm{off}}(A)>P(A)+\CalDelta{}$ (Section~\ref{sec:channels}). Passing calibration means
non-rejection. Detection compares the engine arms directly.

\label{sec:exercise}
\paragraph{Checking speculative execution}
Pass-through drafter wrappers count proposals and verification steps; zero proposals make an arm
\emph{inconclusive}. On all \MOneCPass{} development configurations, a preregistered check found identical outputs, call counts and state digests with and without wrappers, but excluded unobservable llama-cpp-python random-number state, log-probabilities and consecutive draws.

\subsection{Channels S and R}
\label{sec:channels}

\paragraph{Structural witnesses} A \emph{structural witness} is a speculation-on outcome with $P=0$ that violates a registered predicate from documented behavior
(e.g., allowed-token membership), while no speculation-off outcome violates that predicate. A $P=0$ outcome without a
predicate violation is a numerical zero, never a witness.

\paragraph{Differential bounds (channel S)} Before any draw, \badc{} fixes a finite, complement-closed event family
$\mathcal{A}$: one singleton per outcome in the support of $P$ and of the registered hypotheses, an ``other'' event,
token prefixes of length one and two, output lengths, termination codes, a crash event, the contract-violation
event, and one event $A_H$ per registered hypothesis (Section~\ref{sec:hypotheses}). With one-sided
Clopper--Pearson bounds $L$ and $U$~\cite{clopper1934}, it computes
$D=\max_{A\in\mathcal{A}}\bigl(L_{\mathrm{on}}(A)-U_{\mathrm{off}}(A)\bigr)$. Each bound has level $\alpha/(mB)$
for $m$ configurations and $B$ bounds per configuration. Under within-arm i.i.d.\ Bernoulli event
indicators, all channel-S, channel-R and calibration bounds hold jointly with probability at least
$1-\alpha$, without requiring between-arm or between-event independence. Seed IDs do not establish
i.i.d.\ sampling. Compatibility and confirmation have separate budgets. Let $\mu_{\mathrm{off}}$ be the law of the engine's speculation-off arm. Since
$\mathrm{TV}(\mu,\mu_{\mathrm{off}})\geq\mu(A)-\mu_{\mathrm{off}}(A)$ for every $A$, $D$ lower-bounds the total-variation
distance between the arms with that confidence. Channel~S fails if a witness exists or $D>\TvThreshold{}$. Crashes are
kept as outcomes but do not count as silent corruption, so evidence is \emph{eligible} (E4, Section~\ref{sec:confirm}) only
if it is a witness or a bound above \TvThreshold{} on a crash-free event.

\paragraph{Records (channel R)} Where the engine returns per-token records, \badc{} compares them with the
documented record function $r_E$ from teacher-forced replay; a record failure needs
$L_{\mathrm{on}}-U_{\mathrm{off}}>0$ for the any-mismatch event or a per-position event. Records without an
identifiable reference make channel~R incomplete, with no record inference.

\subsection{Testing Supplied Mechanisms with Executable Hypotheses}
\label{sec:hypotheses}

For deterministic prompt-lookup drafters, an executable hypothesis models a suspected faulty schedule
of drafter/processor calls, state updates and loop exit. It enumerates the resulting law $\mu_H$
from the same kernel. Oracle replay follows sequential decoding; repair confirmation tests a patched engine. H-D1-phys, for instance,
applies the processor to the evaluation buffer while taking logits at the logical prefix. Transformers' draft-model path (RQ3 and Section~\ref{sec:discussion}) is outside
Lemma~\ref{lem:composition}; there $\mu_H$ only defines events, and no RQ3 decision needs it to be exact.

Registered hypotheses cover the LCP buffer/stopping schedules and Transformers' guidance, SynthID
and grammar-processor calls. The supplement gives their identifiers and simulator checks.

Each hypothesis adds $A_H=\{y:\mu_H(y)>P(y)\}$ to $\mathcal{A}$; its support singletons stay when RQ3 withholds
$A_H$ and its complement. $H$ is \emph{compatible} if every event's two-sided interval for the speculation-on arm
contains $\mu_H(A)$ within an allowance of \CalDelta{} and no observed outcome has $\mu_H=0$. Each configuration registers at most one defect hypothesis. Compatibility makes a failure
\emph{consistent with its registered hypothesis}; we also report whether the null law $P$ is rejected. This check neither discriminates
between competing mechanisms nor enters a detection decision.

\subsection{Confirmation and Eligibility}
\label{sec:confirm}

\input{tables/assurance-scope}

A \emph{frozen repair} is a minimal patch at the hypothesized site committed before confirmation data.
A cause counts only if fresh speculation-on draws from the repaired engine bound the remaining discrepancy on frozen events. Repairs may not disable speculation, bypass the processor
or change the returned-output contract. The llama-cpp-python repair passes the logical prefix to processors; the
Transformers repair re-creates the stateful processor at each call and replays the prefix it is given, a mechanism
intervention, not a deployment fix. The frozen events qualified in discovery and are listed in Table~\ref{tab:s3}; they have their own error budget. Confirmation requires exercised speculative arms, a fresh
speculation-off arm calibrated over the full event family, the discovery-stage consistency, recurrence
($L_{\mathrm{on}}-U_{\mathrm{off}}>\TvThreshold{}$ on a frozen crash-free event, or fresh witnesses), and simultaneous
upper bounds of at most \RepairGap{} on the repaired arm's gap to the speculation-off arm and $\RepairRate{}$ on its
witness rate. One attempt is allowed per cause and build; a registered priority picks among configurations sharing
a cause.

Counting also requires a documented contract met by the baseline, calibration, observable speculation,
eligible evidence, source-linked confirmation and dated searches excluding public reports or documented
limitations. Causes are deduplicated by code site across versions and manifestations and assigned to an
implementation family. The supplement lists (E1)--(E9). Record-only, crash-only and unresolved-prior-status
findings do not count (Table~\ref{tab:assurance}).

\label{sec:protocol}
\paragraph{Protocol discipline}
Main-study configurations, seeds, events, thresholds, hypotheses, repairs and decision rules were sealed
with harness hashes before confirmatory draws, after source reading and a disclosed pilot. Run scripts check
seals, images and completeness; an audit re-derived decisions (Section~\ref{sec:threats}). Later studies have separate seals.

%% file: tables/assurance-scope.tex
\begin{table}[t]
\caption{Assurance boundaries for prompt lookup. Calibration is non-rejection under the adapter and numerical assumptions.}
\label{tab:assurance}
\centering
\footnotesize
\begin{tabularx}{\columnwidth}{@{}p{0.22\columnwidth}X@{}}
\toprule
Check & Supported interpretation \\
\midrule
Detection & A contract witness or a confidence lower bound on the arms' total-variation distance. Non-rejection is not conformance. \\
Compatibility & Consistency with one supplied mechanism law under the registered rule; alternatives are not excluded. \\
Repair & Fresh recurrence and upper bounds on frozen events, not a global conformance guarantee. \\
Exercise & Hooked draft proposals. The LCP repair acceptance counters are inexact; they do not establish precise acceptance or speedup. \\
\bottomrule
\end{tabularx}
\end{table}

%% file: sections/05-design.tex
\section{Study Design}
\label{sec:design}

\paragraph{Engines and builds}
The main study tests prompt lookup in Hugging Face Transformers' \texttt{generate()}~\cite{wolf2020transformers,sw-transformers}
and llama-cpp-python's Python loop~\cite{sw-llamacpppython,sw-llamacpp}.
Transformers builds are 5.17.0, 5.17.0 with xgrammar 0.2.8~\cite{dong2024xgrammar}, and 5.12.0 for one public control;
llama-cpp-python builds are 0.3.19 and 0.3.35, from published sources.
All \NumBuilds{} ran on one CPU in network-disabled containers pinned by image digest.
We abbreviate the engines as HF and LCP.

\paragraph{Models and prompts}
Fixed seeds generate locally enumerable tiny random networks without third-party weights:
a one-layer Llama-style network for Transformers ($V=\VocabHF{}$, $L=\LHF{}$), and a GGUF network for
llama-cpp-python ($V=\VocabLCP{}$, $L=\LLCP{}$). Both use \texttt{abcdabcdab}, whose repeated $n$-grams
encourage drafting. Matches of length at most \NgramHF{} propose up to \DraftHF{} (HF) or \DraftLCP{} (LCP) tokens.

\paragraph{Configurations}
Stage S1 contains \NumControls{} controls among \NumConfigs{} configurations (Table~\ref{tab:s12}):
\NumNegControlsW{} plain-sampling negative controls and \NumPosControlsW{} public Transformers defects.
The latter are a stop string completed within a verification block (unverified on 5.18.0, which changed stop handling;~\citealp{sw-hf-issue48039}) and a 5.12.0 length overshoot absent from 5.17.0.
The \NumProbes{} S2 probes target Section~\ref{sec:problem}'s classes.
Source reading and a disclosed pilot informed selection (Section~\ref{sec:protocol}); these are tests of
supplied mechanisms and frozen repairs, rather than blind discovery.

\paragraph{Statistics and stages}
Each S1/S2 arm uses \NperArm{} draws; both stages share a multiplicity family ($m=\Mstage{}$, $\alpha=0.05$).
Failed calibration (A5) voids a configuration and bars its build from confirmation.
S3 confirms eligible S2 failures whose causes were not classified as public before sealing, selecting one
configuration per cause/build by Section~\ref{sec:rq2}'s priorities. Confirmation uses fresh seeds,
\NSThreeLCPTwoOneNine{} draws per unrepaired arm and a separate budget ($m=\SThreeM{}$;
Section~\ref{sec:confirm}). S1--S3 used \TotalDraws{} million decision draws and \CoreHours{} CPU core-hours.

\paragraph{Held-out transfer (RQ2)}
A separate sealed study repeats \NumConfirmW{} confirmed probes on \MOneBK{} held-out tiny random model/prompt
variants each, without retuning; only model seeds and prompts change ($m=18$ in discovery, $15$ in confirmation).

\paragraph{Additional mechanism and output studies (RQ2)}
Separate freezes cover the returned-output marginals from S3, processor scores and controls on
SmolLM2-135M-Instruct~\cite{allal2025smollm2}, and one pilot-selected returned-token event.
Historical decisions remain unchanged (Supplement Sections D--G).

\paragraph{Controlled comparison (RQ3)}
A preregistered comparison varies configuration selection (directed/random) and inclusion of
$A_H$ and its complement, sharing draws within each of \MFourCampaigns{} fresh-model/prompt campaigns.
A cause is recovered when fresh draws confirm the frozen repair for its higher-priority detected configuration.
The primary contrast compares recovered causes under directed and random selection with $A_H$.
A permutation chi-square comparator (C) uses full outputs without an oracle or witnesses. Shared feature families, ranges,
prompts and repairs limit this to selection given that knowledge; the supplement gives the engine/path matrix.

\paragraph{Workflow comparison}
A separate registered workflow comparison uses two LLM sessions per arm on screened Transformers fixes,
with the same time and token budgets. Section~\ref{sec:discussion} gives its limited inference;
Supplement Section J retains selection, budgets and deviations.

\paragraph{Transfer to vLLM (RQ4)}
The unchanged channel-S rules and thresholds, with a vLLM-specific event family, test vLLM 0.8.5.post1~\cite{kwon2023vllm,sw-vllm}
with $n$-gram drafting on L4 GPUs. Each of \MThreeCampaigns{} campaigns uses one held-out tiny random Qwen2 model
($V=8$, $L=3$). The reference enumerates the speculation-off law from CPU 32-bit floating-point (fp32) Transformers Qwen2 logits with a
NumPy sampler mirror; the engine uses 16-bit floating point (fp16). Probability-gap bounds compare engine arms directly; the reference
supplies events and validation gates. Passing calibration means failure to reject, not numerical identity.

On vLLM's V1 engine we test five public families: allowed tokens, bad words, logit bias, minimum tokens and stop strings; on the legacy V0 engine we test allowed-token processor rows. The supplement retains their source references,
fix/refusal status and the host-runtime constraints on newer-build execution.

Each family has a directed probe and one random configuration per campaign. Controls include plain
sampling, temperature/top-$k$/top-$p$ row-path checks and V1 fallback checks; exposure is labeled independently of the verdict (supplement). Channel~S omits hypothesis events ($m=20$ per campaign,
200{,}000 draws per arm); directed detections are repeated on fresh draws. Public families and no repair
arm make this a transfer test, not a search for new defects.

%% file: sections/06-results.tex
% Results: RQ1-RQ2 (06a) and RQ3-RQ4 (06b)
\input{sections/06a-results}

\input{sections/06b-results}

%% file: sections/06a-results.tex
\section{Results}
\label{sec:results}

\subsection{RQ1: Calibration and Detection}
\label{sec:rq1}

\input{tables/tab-s12}

All \NumControls{} controls behaved as registered (Table~\ref{tab:s12}, stage S1), and every S2 configuration
passed calibration, so none was voided. The \NumNegControlsW{} negative controls were exercised (HF17-N0 proposed
\PropHFOneSevenNZero{} draft tokens and accepted \AccHFNZero{}) and not rejected. Both public Transformers defects were detected through structural witnesses: generation continued past a stop string completed inside a
verification block (HF17-P1), and release 5.12.0 returned more than $L$ tokens (HF12-P2); their differential bounds
stayed below the threshold.

\NumEligibleFailWCap{} probes failed channel~S with eligible evidence (Table~\ref{tab:s12}).

\paragraph{llama-cpp-python, prefix-dependent processors}
With a harness processor that allows different token sets at even and odd absolute positions, the speculation-on arm
emitted a forbidden token in \WitLCPOneOneNine{} of \NperArm{} draws on 0.3.19 and \WitLCPOneThreeFive{} on 0.3.35,
against none with speculation off ($D=\DLCPOneOneNine{}$ and \DLCPOneThreeFive{}). With the library's
\texttt{MinTokensLogitsProcessor} and a minimum of \MinTokensM{} tokens, the law of returned completions shifted by
$D=\DLCPTwoOneNine{}$ (0.3.19) and \DLCPTwoThreeFive{} (0.3.35). No minimum-length predicate is registered, because
the completion API hides some internal tokens and a short visible completion can be legitimate, so the evidence is
the bound alone.

\paragraph{Transformers, stateful processors}
The SynthID watermark shifted the law by $D=\DHFOneSevenSeven{}$, and classifier-free guidance
by \DHFOneSevenSix{}, just above the threshold. Both built-in processors keep state across calls. Guidance maintains its own unconditional context
and cache; SynthID maintains an $n$-gram context and a call counter. The
third-party xgrammar 0.2.8 processor also failed ($D=\DHFXEight{}$), but crashes dominate it: \ErrHFXEight{} of the \NperArm{} speculative generations raised an
assertion inside the grammar matcher, and on crash-free events the bound is only \DnHFXEight{}, so we report a
robustness finding, not silent corruption.

\paragraph{State trace and intervention}
Prompt lookup can call stateful processors before verification; discarding drafts or cropping the target
cache leaves their updates. The supplement gives the source-derived schematic; Section~\ref{sec:trained} reports the trained-checkpoint trace.
The intervention preserves chain order and rebuilds sequential prefix state on every call.
Earlier guidance and SynthID replay calls can use zero scores because state updates ignore them;
the final call uses supplied scores. Assurance remains the frozen-event bound (Section~\ref{sec:rq2}),
not a deployment guarantee.

\paragraph{Not rejected}
With top-$k$, top-$p$ and repetition penalty together (HF17-3) the drafter proposed no token, so speculation was
silently disabled and the configuration is inconclusive, not passing. The remaining probes
were not rejected (Table~\ref{tab:s12}; llama-cpp-python's record probe only on returned outputs, since its
channel~R is incomplete). These non-findings are not evidence of
conformance: power over the event family was not evaluated. The supplement gives a post hoc single-event scale heuristic.

\subsection{RQ2: Mechanism Compatibility, Repair Confirmation and Transfer}
\label{sec:rq2}

\input{tables/tab-s3}

All \NumEligibleFailW{} eligible failures rejected the null and were compatible with their registered
mechanism (Table~\ref{tab:s12}). The supplementary appendices give exact mechanism distances and a
post hoc check that removes hypothesis events. RQ3 tests the effects of those events and test selection
in a controlled comparison.

Stage S3 confirmed one configuration per cause and build (Table~\ref{tab:s3}). On fresh seeds the unrepaired engines
failed again on their frozen events (recurrence \RecHFOneSevenSix{}, \RecLCPTwoOneNine{} and \RecLCPTwoThreeFive{}).
With speculation still exercised, the simultaneous upper bounds on the repaired arms' remaining gap to the
speculation-off arm were \GapHFOneSevenSix{}, \GapLCPTwoOneNine{} and \GapLCPTwoThreeFive{}. SynthID's failure is confirmed only through its shared
cause: the priority sealed with each configuration selected guidance (HF17-6) over SynthID (HF17-7), as it selected
the minimum-length probe (LCP-2) over the harness processor (LCP-1), and SynthID was not repaired separately.

\paragraph{Counting causes}
The two llama-cpp-python manifestations share one code site across both builds, and guidance and SynthID share the
missing rollback of processor state, so each pair counts as one cause. A stopping
re-check that sees unrolled drafts (LCP-3) was compatible with its hypothesis, but its effect ($D=\DLCPThreeOneNine{}$;
exact distance \TvHLCPThreeOneNine{}) stayed below the threshold, so it does not count. \NumCausesWCap{}
eligible causes remain, one per implementation family.

\paragraph{Prior reports and source persistence}
Dated searches (\QueriesA{} and \QueriesB{} queries, logged before confirmation data) found no report of
either mechanism. Later snapshots retained the sites (supplement), without historical execution or repair
coverage. The score pilot below executes guidance in 5.20.2.

\paragraph{Held-out models and prompts}
On the held-out models and prompts (Section~\ref{sec:design}), the minimum-length probe was detected and
repair-confirmed in \MOneBLCPNineteenDet{} of \MOneBK{} variants on 0.3.19 and \MOneBLCPThirtyFiveDet{} of \MOneBK{}
on 0.3.35. The guidance probe was detected in \MOneBHFGuidDet{} of \MOneBK{}, exceeding our prediction of at most three, and all \MOneBHFGuidAtt{} confirmations attempted under the preregistered cap were confirmed. This covers tiny
random models with one horizon and prompt length per engine.

\paragraph{Which returned-output marginals change?}
A retrospective, nominal simultaneous 95\% construction covers every union of categories for each of five observables
(Supplement Section D). In the LCP minimum-length case, faulty/off length TV is 14.1202\% versus 0.2686\%
for the first token: the length lower bound, 13.7314\%, exceeds the first-token upper bound, 0.7162\%.
Intervention/off length and termination upper bounds fall below 1\%, but the first-token bound is 1.0355\%.
These binomial-assumption-dependent bounds establish neither all-marginal nor full-sequence conformance.

\paragraph{Trained-checkpoint mechanism studies}
\label{sec:trained}
Across four prompts in Transformers 5.17.0/5.20.2, all 64 speculative requests per release had a committed-prefix
history mismatch and conditional-score TV above 0.001 (Supplement Section E).
A frozen eight-prompt 5.20.2 matrix exercised plain and repetition-penalty controls that
agreed with source-defined target-score references within $10^{-4}$. Native CFG had witnesses at 6/8 fixed
initial states (12/16 planned requests), including all four lookup-oriented states; P05/P07 were unexposed.
Functional prefix-history controls agreed with references while changing history and cache execution (Supplement Section G).
Repeated initial states are not independent; neither score study tests returned-output laws or prevalence.

\paragraph{Returned-output event}
A frozen follow-up at the pilot-selected Paris factual-sentence prompt used the same checkpoint and Transformers 5.20.2.
The event was whether the first returned token belonged to a fixed set of 1,021 token IDs. We fixed this
set from pilot scores before collecting 1,024 fresh requests per arm. Event rates were 80.37\% for
ordinary decoding, 14.16\% for native prompt lookup and 81.35\% for the functional prefix-history control.
Under within-arm i.i.d.\ Bernoulli sampling, nominal simultaneous 95\% bounds lower-bound the
ordinary-minus-native event gap (and returned-sequence TV) by 60.26 points and upper-bound the
absolute control/ordinary gap by 6.95 points. Supplement Section F retains counts, shared allocation,
event-specific control scope and the output-directory relocation.
A second-host run reused the collector, checkpoint and seeds: all 3,072 first returned tokens and
event counts matched. This is reproducibility evidence for the same implementation on one additional
host, without independent sampling or narrower intervals (Supplement Section H).

\input{tables/tab-evidence-scope}

%% file: tables/tab-s12.tex
\begin{table*}[t]
\caption{The \NumConfigs{} sealed S1 controls and S2 probes, \NperArm{} draws/arm. HF17/HF12: Transformers
5.17.0/5.12.0; HFX: HF17 with xgrammar 0.2.8; LCP suffixes: 0.3.19/0.3.35.
NR: not rejected; FAIL\textsubscript{S}: channel-S failure; INC.: inconclusive. Witness: speculative draws
with a structural witness. $D$: simultaneous lower bound (Section~\ref{sec:channels}); negative values mean that no event separates the arms. Eligible: witness or crash-free bound above \TvThreshold{}. All listed hypotheses (H- prefix omitted) are compatible.
Proposed: speculative draft-token count.}
\label{tab:s12}
\centering
\scriptsize
\setlength{\tabcolsep}{4pt}
\begin{tabular}{@{}llllrrllr@{}}
\toprule
Config & Feature probed & Role & Decision & Witness & $D$ & Eligible & Hypothesis & Proposed\\
\midrule
\input{tables/s12.tex} 
\bottomrule
\end{tabular}
\end{table*}

%% file: tables/s12.tex
% generated by paper/scripts/make_tables.py from harness/configs/m0 and results/m0/S1,S2 -- do not edit
HF12-P2 & plain sampling (5.12.0) & pos. control & FAIL\textsubscript{S} & 623 & 0.0025 & yes & -- & 925{,}878 \\
HF17-N0 & plain sampling ($T{=}1$) & neg. control & NR & 0 & \ensuremath{-}0.0001 & no & -- & 927{,}150 \\
HF17-P1 & \texttt{stop\_strings}=["cd"] & pos. control & FAIL\textsubscript{S} & 1{,}144 & 0.0052 & yes & -- & 925{,}554 \\
LCP-N0-19/35 & plain, \texttt{generate()} & neg. control & NR & 0 & \ensuremath{-}0.0001 & no & -- & 1{,}067{,}543/1{,}067{,}406 \\
LCP-N1-19/35 & plain, \texttt{create\_completion()} & neg. control & NR & 0 & \ensuremath{-}0.0001 & no & -- & 985{,}359/984{,}606 \\
HF17-3 & $T{=}0.7$, top-$k$ 3, top-$p$ 0.9, rep.\ 1.3 & probe & INC. & 0 & \ensuremath{-}0.0001 & no & -- & 0 \\
HF17-4 & \texttt{min\_new\_tokens}=2 & probe & NR & 0 & \ensuremath{-}0.0001 & no & -- & 969{,}402 \\
HF17-5 & sequence bias on one token & probe & NR & 0 & \ensuremath{-}0.0001 & no & -- & 776{,}481 \\
HF17-6 & \texttt{guidance\_scale}=1.5 (CFG) & probe & FAIL\textsubscript{S} & 0 & 0.0142 & yes & D2-cfg & 933{,}579 \\
HF17-7 & SynthID watermark & probe & FAIL\textsubscript{S} & 0 & 0.6417 & yes & D2-synthid & 713{,}529 \\
HF17-9 & plain + \texttt{output\_scores} & probe (records) & NR & 0 & \ensuremath{-}0.0001 & no & -- & 927{,}516 \\
HFX-8 & xgrammar grammar (third party) & probe & FAIL\textsubscript{S} & 0 & 0.4060 & no & D2-xgr & 654{,}390 \\
LCP-1-19 & prefix-parity processor & probe & FAIL\textsubscript{S} & 23{,}335 & 0.1131 & yes & D1-phys & 1{,}145{,}510 \\
LCP-1-35 & prefix-parity processor & probe & FAIL\textsubscript{S} & 23{,}407 & 0.1135 & yes & D1-phys & 1{,}145{,}054 \\
LCP-2-19 & \texttt{MinTokensLogitsProcessor}(3) & probe & FAIL\textsubscript{S} & 0 & 0.1490 & yes & D1-phys & 1{,}145{,}577 \\
LCP-2-35 & \texttt{MinTokensLogitsProcessor}(3) & probe & FAIL\textsubscript{S} & 0 & 0.1475 & yes & D1-phys & 1{,}143{,}741 \\
LCP-3-19/35 & stopping criterion (last id) & probe & NR & 0 & 0.0078/0.0076 & no & D3-postcheck & 901{,}655/900{,}158 \\
LCP-4-19/35 & \texttt{stop}=["da"] & probe & NR & 0 & \ensuremath{-}0.0001 & no & -- & 968{,}382/968{,}624 \\
LCP-5-19/35 & \texttt{logprobs}=10 & probe (records) & NR & 0 & \ensuremath{-}0.0001 & no & -- & 984{,}488/984{,}393 \\
LCP-6-19 & repeat/freq./presence penalties & probe & NR & 0 & \ensuremath{-}0.0001 & no & -- & 996{,}477 \\
LCP-7-19 & GBNF grammar & probe & NR & 0 & \ensuremath{-}0.0001 & no & -- & 1{,}447{,}957 \\

%% file: tables/tab-s3.tex
\begin{table*}[t]
\caption{Stage S3 confirmation on fresh seeds (repair arm
\NRepHFOneSevenSix{} draws for Transformers and \NRepLCPTwoOneNine{} for llama-cpp-python). Recurrence:
$L_{\mathrm{on}}-U_{\mathrm{off}}$ on the event frozen in S2 (above \TvThreshold{}). Repair gap: simultaneous
upper bound on the repaired arm's gap to the speculation-off arm on that event only (at most
\RepairGap{}). Witness UB: the same for its witness rate (at most $\RepairRate{}$). $\Delta B$:
speculation-induced bias (Section~\ref{sec:discussion}), sealed Hoeffding interval at level $1-0.05/\SThreeM{}$ per
configuration (95\% jointly).}
\label{tab:s3}
\centering
\scriptsize
\setlength{\tabcolsep}{4pt}
\begin{tabular}{@{}llllrrll@{}}
\toprule
Config & Build & Frozen event & Recurrence & Repair gap & Witness UB & Confirmed & $\Delta B$ [interval]\\
\midrule
\input{tables/s3.tex} 
\bottomrule
\end{tabular}
\end{table*}

%% file: tables/s3.tex
% generated by paper/scripts/make_tables.py from results/m0/S3 -- do not edit
HF17-6 & 5.17.0 & $A_{\text{D2-cfg}}$ & 0.0195 & 0.0032 & $6.8\times 10^{-6}$ & yes & $\ensuremath{-}0.0032$ [$\ensuremath{-}0.0132$, $0.0067$] \\
LCP-2-19 & 0.3.19 & $A_{\text{D1-phys}}$ & 0.1549 & 0.0020 & $3.4\times 10^{-5}$ & yes & $\ensuremath{-}0.0047$ [$\ensuremath{-}0.0146$, $0.0053$] \\
LCP-2-35 & 0.3.35 & $A_{\text{D1-phys}}$ & 0.1548 & 0.0023 & $3.4\times 10^{-5}$ & yes & $\ensuremath{-}0.0042$ [$\ensuremath{-}0.0142$, $0.0057$] \\

%% file: tables/tab-evidence-scope.tex
\begin{table}[t]
\caption{Evidence units and limits of the follow-ups. Repeated calls or seeds do not add independent settings;
none estimates workload prevalence.}
\label{tab:evidence-scope}
\centering
\small
\begin{tabular}{@{}p{0.23\columnwidth}p{0.70\columnwidth}@{}}
\toprule
Evidence & Claim and limit \\
\midrule
Historical repairs & Selected tiny-model probes; fresh-draw bounds on frozen repair events. \\
Trained scores & Four-prompt pilot and eight fixed matrix states; conditional-score/history checks, not returned-law inference. \\
Returned event & One selected prompt/checkpoint, 1,024 requests/arm; nominal event-gap bound under within-arm i.i.d.\ sampling. \\
Second host & Same collector, prompt, event and seeds; reproducibility without pooled $N$. \\
\bottomrule
\end{tabular}
\end{table}

%% file: sections/06b-results.tex
\subsection{RQ3: Controlled Comparison}
\label{sec:rq3}

\input{tables/tab-rq3}

All 30 controls and all 10 repaired controls (Section~\ref{sec:design}) behaved as registered,
and both causes were recovered in nearly every campaign and cell (Table~\ref{tab:rq3}a). With $A_H$, directed and random selection
recovered the same number of causes in \MFourdAZeros{} of \MFourCampaigns{} campaigns, so their difference $d_A$ is
zero wherever it is determined. Campaign~3 is only bounded: one directed confirmation driver failed before producing
raw outputs, and its unit was recorded INVALID (Section~\ref{sec:threats}). The exact 95\% upper bound on the
probability that $d_A$ is nonzero in a campaign is \MFourdACP{}. Removing $A_H$ changed the recovered causes in at most one
campaign per selection policy ($d_B=0$ in \MFourdBDZeros{} directed and \MFourdBUZeros{} random campaigns). Both
registered predictions on $d_A$ and $d_B$ hold.

Including the hypothesis event mainly affected detection of the guidance family (Table~\ref{tab:rq3}b). With $A_H$, \badc{} detected
guidance in 3 directed and 5 random configurations of 10, and without it in 1 and 2; the other families were
unaffected. Each cause has a second family that \badc{} detected in at least 9 campaigns per policy (SynthID for the stateful-processor cause, and minimum length or prefix parity for the processor buffer), so missed guidance detections affected cause recovery once: in campaign~2 the random SynthID configuration was not detected, and only $A_H$ made the random guidance configuration eligible ($d_B=1$).

The permutation comparator~C (Section~\ref{sec:design})
detected guidance in 10 directed and 9 random configurations and every other family at least as often as \badc{}, so
our registered prediction that C would detect guidance less often than \badc{} with $A_H$ failed. The two tests answer
different questions: C rejects equality of the two laws at level $0.05/11$, whereas channel~S needs a structural
witness or a simultaneous lower bound above \TvThreshold{} on their total-variation distance. The study used
\MFourCoreK{} thousand core-seconds.

These comparisons establish neither equivalence nor detection advantage. Table~\ref{tab:assurance}
summarizes the additional reference, mechanism and frozen-repair evidence.

\subsection{RQ4: Transfer to vLLM}
\label{sec:rq4}

\input{tables/tab-rq4}

\paragraph{Positive families}
Of the \MThreePlanned{} planned positive configurations, \MThreeMissingDisc{} were lost to failures of our GPU wrapper
(recorded INVALID, not re-run) and \MThreeUnexposed{} random ones gave their family no opportunity. \badc{} and the
permutation comparator each detected all \MThreeExposed{} exposed configurations (\MThreeExposedD{} directed,
\MThreeExposedU{} random; Table~\ref{tab:rq4}a). All
families but logit bias were detected through structural witnesses of their predicate; logit bias, which registers no
family-specific predicate, was detected through its bounds. Of the \MThreeRedetReq{} directed detections, \MThreeRedet{} recurred on fresh draws; the remaining confirmation was lost to a wrapper failure. Small-cell intervals and missing-unit bounds are in the supplement. The yield is conditional on exposure. The five V1 families
are defect classes, not five independent causes: in all but stop strings, 0.8.5.post1's verification of draft rows ignores a
sampling parameter.

\paragraph{Controls}
On V1, no negative control or row-path check was rejected in any campaign, and all \MThreeFallbackN{} fallback
checks proposed no draft token and were reported as inconclusive, not passing, as predicted. On V0, plain
sampling with per-request seeds was rejected in all \MThreeVZeroCtlRej{} campaigns (\MThreeVZeroCtlDet{} with eligible
evidence) and the row-path check in \MThreeVZeroTtpRej{}, with outputs outside the top-$k$/top-$p$ support, so the registered prediction of clean controls
fails. Post hoc, on campaign~1's model, a two-token
output had probability 0.024 without and 0.069 with speculation for seeded requests (0.025 and 0.024
unseeded). V0's seeded noise initialization and generator indexing also make seeded draws dependent, so the V0 bounds and
permutation p-values lack their nominal guarantees. V0's speculative decoding was removed upstream in 0.10.0; a sealed repair study of the seeded V0 path stopped incomplete, so we
count no V0 defect. Two registered predictions on per-family detection are not determined because a bad-words unit is
missing. The study used \MThreeGPUh{} GPU-hours.

Thus the unchanged rules detected all 45 valid, exposed V1 configurations with clean controls.
Source reading finds four public causes fixed or refused in 0.30.0; stop strings remain runtime-unverified there.
Failed V0 controls leave its 7 exposed detections (and \MThreeUnexposedDet{} unexposed ones) unattributed,
likely due to seeded speculative sampling. All families have public causes; RQ4 adds no counted defect.

%% file: tables/tab-rq3.tex
\begin{table}[t]
\caption{RQ3, \MFourCampaigns{} campaigns on held-out models. (a) Campaigns in which each cause was recovered (repair-confirmed on
fresh draws) and total recovered causes, per selection policy and event family; a range is a bound over missing units.
(b) Configurations detected, of 10 per family and policy, by channel~S with and without the hypothesis event $A_H$ and
by the permutation comparator~C.}
\label{tab:rq3}
\centering
\scriptsize
\setlength{\tabcolsep}{3pt}
\begin{tabularx}{\linewidth}{@{}Xccc@{}}
\toprule
(a) Selection, events & stateful processors & processor buffer & causes (of 20) \\
\midrule
\input{tables/rq7a.tex} 
\bottomrule
\end{tabularx}

\smallskip
\begin{tabularx}{\linewidth}{@{}Xcccccc@{}}
\toprule
& \multicolumn{3}{c}{directed} & \multicolumn{3}{c}{random} \\
\cmidrule(lr){2-4}\cmidrule(l){5-7}
(b) Family & with $A_H$ & without & C & with $A_H$ & without & C \\
\midrule
\input{tables/rq7b.tex} 
\bottomrule
\end{tabularx}
\end{table}

%% file: tables/rq7a.tex
%% generated by paper/scripts/make_m4_tables.py from results/m4/ANALYSIS.json; do not edit
directed, with $A_H$ & 10 & 9--10 & 19--20 \\
directed, without $A_H$ & 10 & 9--10 & 19--20 \\
random, with $A_H$ & 10 & 10 & 20 \\
random, without $A_H$ & 9 & 10 & 19 \\

%% file: tables/rq7b.tex
%% generated by paper/scripts/make_m4_tables.py from results/m4/ANALYSIS.json; do not edit
guidance & 3 & 1 & 10 & 5 & 2 & 9 \\
SynthID & 10 & 10 & 10 & 9 & 9 & 10 \\
minimum length & 10 & 10 & 10 & 6 & 6 & 6 \\
prefix parity & 7 & 7 & 7 & 10 & 10 & 10 \\

%% file: tables/tab-rq4.tex
\begin{table}[t]
\caption{RQ4, vLLM 0.8.5.post1 on five held-out tiny models. (a) Positive families: configurations whose speculative
arm gave the defect an opportunity (exp.), detected by \badc{} and by the permutation comparator~C, and directed
detections that recurred on fresh draws; random configurations also list the valid ones (one planned per campaign).
The V1 rows sum to 45 exposed configurations (24 directed, 21 random), the V0 row to 7 (5 and 2). The V0 row is
unattributed (V0's negative controls failed), and its bounds and p-values lack their nominal guarantees. (b)
Controls rejected by
\badc{} per path, and fallback checks whose speculative arm proposed a draft token.}
\label{tab:rq4}
\centering
\scriptsize
\setlength{\tabcolsep}{2.4pt}
\begin{tabularx}{\linewidth}{@{}Xlcccccccc@{}}
\toprule
& & \multicolumn{4}{c}{directed} & \multicolumn{4}{c}{random} \\
\cmidrule(lr){3-6}\cmidrule(l){7-10}
(a) Family & path & exp. & \badc{} & C & recurred & valid & exp. & \badc{} & C \\
\midrule
\input{tables/rq8a.tex} 
\bottomrule
\end{tabularx}

\smallskip
\begin{tabularx}{\linewidth}{@{}Xcc@{}}
\toprule
(b) Check & V1 & V0 \\
\midrule
\input{tables/rq8b.tex} 
\bottomrule
\end{tabularx}
\end{table}

%% file: tables/rq8a.tex
%% generated by paper/scripts/make_m3_tables.py from results/m3/ANALYSIS.json; do not edit
allowed tokens & V1 & 5 & 5 & 5 & 5 of 5 & 4 & 4 & 4 & 4 \\
bad words & V1 & 4 & 4 & 4 & 3 of 4 (1 missing) & 5 & 4 & 4 & 4 \\
minimum tokens & V1 & 5 & 5 & 5 & 5 of 5 & 5 & 5 & 5 & 5 \\
stop strings & V1 & 5 & 5 & 5 & 5 of 5 & 5 & 3 & 3 & 3 \\
logit bias & V1 & 5 & 5 & 5 & 5 of 5 & 5 & 5 & 5 & 5 \\
allowed tokens & V0 & 5 & 5 & 5 & 5 of 5 & 5 & 2 & 2 & 2 \\

%% file: tables/rq8b.tex
%% generated by paper/scripts/make_m3_tables.py from results/m3/ANALYSIS.json; do not edit
negative control (plain sampling) & 0 of 5 & 5 of 5 \\
row-path check ($T=0.7$, top-$k$ 5, top-$p$ 0.9) & 0 of 5 & 4 of 5 \\
fallback checks (drafting off per request) & 0 of 20 & -- \\

%% file: sections/07-discussion.tex
\section{Discussion and Threats to Validity}
\label{sec:discussion}

\paragraph{Processor-interface implications}
The two causes show why correct model logits are insufficient. A llama-cpp-python callback reads
unverified drafts from the evaluation buffer; Transformers' candidate filtering advances processor state
before verification. The frozen interventions support these mechanisms.

For the standard guarantee, a prefix-sensitive callback needs a position-indexed view, and a stateful
callback needs the sequential state associated with that view. This includes effects of candidate
filtering and rejected verification. Cropping the target cache does not establish either condition.
Different state-management strategies can satisfy these obligations.

Trained-checkpoint traces show the guidance-history mismatch already at the first committed target call, before any later rejection. In the control matrix, plain and repetition-penalty controls agree with target-score references, while native guidance has
history witnesses at six fixed states. The functional
guidance control changes history, cache use and execution jointly, so it does not uniquely attribute the
effect or establish a production repair.

\paragraph{What testing can establish}
Returned-law tests quantify observable differences; traces and interventions help interpret them.
A test of the first returned token can miss a change in output length. A conditional-score
anomaly alone need not establish a change in the returned-output distribution. Table~\ref{tab:evidence-scope} separates the follow-ups' units and limits.

\paragraph{Comparison with differential testing}
RQ3 establishes no discovery or cause-recovery advantage. A separate registered workflow comparison had
one eligible, record-only defect and no submitted diagnosis or patch in either arm. Neither \badc{} session ran the oracle: all sessions stopped at their token cap after 1.3--4.7 minutes, so the reported 90-minute porting time is a censoring value. Post hoc shakedown repairs also did not exercise the oracle. The comparison therefore
leaves added diagnostic value open. Supplement Section J retains the full account, bounds and deviations.

\paragraph{What the evidence supports}
Both source-suspected causes have a mechanism law and fresh-draw repair bound; better or faster repairs remain
untested. A diagnostic applies TRL's \texttt{sequence\_mask} importance-weight rule~\cite{sw-trl}, with upper cutoff \TrlCap{}, to a fixed hashed reward $R\in[0,1]$. $\Delta B$ subtracts the arms' weighted mean rewards;
each bias is relative to $\mathbb{E}_P[R]$. All \NumConfirmW{} sealed intervals include zero
(Table~\ref{tab:s3}); these probes return no records, so all weights are one. No trainer--engine path was run.
The replication package also retains a sealed draft-model case study and one record-only failure.

\subsection{Threats to Validity}
\label{sec:threats}

\paragraph{Internal validity}
Oracle and hypothesis laws depend on the adapters and floating-point evaluation. Calibration's non-rejection does not prove exactness or distinguish competing mechanisms.
The vLLM reference uses a CPU fp32 mirror against an fp16 engine (Section~\ref{sec:design}). Two analyses missing
at the main study's post-run audit were added post hoc, changing no decision. Our sealed-document edit (RQ3; the unit whose confirmation driver failed, Section~\ref{sec:rq3}) and GPU-wrapper failures (RQ4) leave affected units INVALID and bounded as missing outcomes.

\paragraph{Construct validity}
The effect threshold (\TvThreshold{}) and event family were sealed before confirmatory draws; they detect only
differences visible on those events. Repair bounds likewise cover frozen events only. Confirmation checks
proposed drafts; review, rather than decision code, checks that repairs keep speculation enabled.

\paragraph{External validity}
Historical distributional results use tiny random networks and \MOneBK{} held-out model/prompt variants
per confirmed probe. One 135M-parameter checkpoint provides a four-prompt score pilot, an eight-prompt
control matrix and one pilot-selected returned-event bound under within-arm i.i.d.\ sampling.
Selection precludes prevalence estimates; natural workloads, long contexts, production-scale models and
wider distributional transfer remain untested. Returned-event evidence includes a recorded output-directory relocation and lacks an independent integrity audit (Supplement Section F).

The historical studies cover pinned CPU builds with prompt lookup (plus one draft-model path) and vLLM 0.8.5.post1
with $n$-gram drafting. They do not cover SGLang or production model-based drafters with dynamic
verification~\cite{deepseekv41flash2026}, or establish that an RL pipeline combines the affected features
with speculation. Source persistence (Section~\ref{sec:rq2}) does not extend runtime or repair coverage;
stop-string behavior on newer Transformers/vLLM builds remains unverified, including after the upstream
token-ID cleanup~\cite{sw-vllm-pr47616}.

A later preregistered L4 run on vLLM 0.30.0 and main ac9126e5 ended inconclusive. Its integrity audit failed
because the candidate bonus-row cause (Supplement Section C.8) is related to a public report~\cite{sw-vllm-pr41768}; the registered rule makes
every claim inconclusive and counts the cause as zero. These records support no current-runtime or
repair-confirmed conclusion.

A separate archived-law simulation illustrates the additional dependence of diagnosis on hypothesis-library
coverage and decision rules, including abstention failures and ambiguous low-sample controls
(Supplement Section I). It is not an end-to-end \badc{} comparison.

Probe selection used source reading and a disclosed pilot that exposed the LCP 0.3.19 violation and both
public Transformers defects, and fixed the Transformers model seed. Random-selection arms inherit this knowledge; RQ3 uses one comparator chosen by us.

%% file: sections/08-related.tex
\section{Related Work}
\label{sec:related}

\paragraph{Speculative decoding guarantees}
Speculative sampling preserves the target distribution~\cite{leviathan2023,chen2023specsampling}, qualified as
``within hardware numerics'' by Chen et al. EAGLE~\cite{li2024eagle} reports the same guarantee; a two-sample
$\chi^2$ test did not distinguish a from-scratch temperature-only decoder from target sampling~\cite{chordiya2026lossless}.

\paragraph{RL rollouts}
NeMo-RL, BubbleSpec and TLT describe lossless acceleration; EfficientRollout retains distribution-preserving
techniques~\cite{nemorl2026,bubblespec2026,adaptivedraft2025,efficientrollout2026}. DeepSeek-V4.1 uses dynamic
verification~\cite{deepseekv41flash2026}; DSpark requires admission independent of unsampled tokens~\cite{dspark2026}.
These works do not compare the composed engine with its speculation-off law; we make no claim about their implementations.

\paragraph{Inference-engine correctness}
A five-engine study of 929 bugs proposes a deterministic-decoding oracle~\cite{liu2026firstlook}; Ekka uses
a reference to diagnose silent errors~\cite{gu2026ekka}; GRIEF fuzzes request traces and checks deterministic
replay~\cite{zhao2026continuous}. These three studies report neither their own speculative-decoding findings
nor sampled-output-law comparisons. Public reports cover greedy mismatch and omitted verification
parameters~\cite{sw-mlc-issue2167,sw-vllm-issue9834}. Other work diagnoses greedy batch-speculation
misalignment~\cite{batchspec2025} or uses speculation as an ``implementation null''~\cite{specnull2026}.

FlashInfer-Bench validates stochastic sampling kernels against expected masked distributions using empirical
total-variation checks and mask predicates~\cite{flashinferbench2026}. We test composed speculative engines'
prefix/state handling, supplied mechanisms and frozen repairs.

\paragraph{Sample verification}
Token-DiFR uses a trusted reference and shared seed; its speculative-verification sketch needs acceptance,
resampling and bonus records~\cite{difr2025}. \badc{} needs neither seed replay nor those records but uses many
tiny-model draws. Cross-engine distance estimation does not vary speculation~\cite{crossengine2026}.

\paragraph{Probabilistic software testing}
Differential testing, probabilistic-system testing and flaky ML tests are established~\cite{mckeeman1998,dutta2018probfuzz,dutta2020flex}.
Equal on/off laws form a metamorphic relation; finite enumeration is a known exact-oracle technique.
\badc{} supplies domain-specific kernel references, witnesses and executable hypotheses for deterministic prompt lookup.

%% file: sections/09-conclusion.tex
\section{Conclusion}

Correct model logits do not suffice when speculative paths expose drafts to processors or retain
nonsequential state. Interventions confirmed two tiny-model causes; trained-checkpoint studies connect
guidance history to one returned-event difference reproduced on a second host with the same seeds.
Assurance remains event-specific. Controlled comparisons establish no diagnostic advantage, historical
vLLM adds no causes, and the archived-law simulation illustrates hypothesis-library dependence.
Workload prevalence, broader model/drafter transfer and RL consequences remain open.

%% file: sections/10-data.tex
\paragraph{Data availability}
\label{sec:data}

A privately prepared replication package binds code, protocols, environment pins and available records, distinguishing originals, exports and withheld logs. Replay reconstructs tables and scoped statistics with numerical differences,
exposure labels and INVALID outcomes retained. Full reproduction under the original seals and independent engine reproduction remain unestablished. The supplement provides the proof and designs; the package will be archived publicly after acceptance.

%% file: sections/11-ai-usage.tex
\section*{AI Usage Disclosure}

Anthropic Claude models drafted experimental and analysis code, tables, the figure specification and
manuscript text. OpenAI GPT models accessed through Codex assisted literature search, protocol and code
review, checks of recorded runs and citations, analysis and manuscript revision. For the marginal-analysis,
trained-checkpoint and archived-law follow-ups, Codex also proposed study designs, implemented collectors,
analyzers and tests, executed local scripts, and drafted interpretations of the results.

Experimental token outputs came from the tested models. Reported numerical results were computed from
recorded data or cited sources. AI-assisted internal reviews do not constitute independent experimental
replication. The authors retain responsibility for the correctness, originality and integrity of the work.

%% file: supplement/arxiv-restatement.tex
\section{Setting (restated from Section 3 of the paper)}
\label{app:setting-restated}
Fix an engine $E$ at a pinned version, a model, a prompt $x$, a generation horizon $L$, and a configuration of
sampling features. The \emph{sequential kernel} $K_E(\cdot\mid h)$ is the processed next-token distribution the
engine would sample from after the committed prefix $h$ in ordinary decoding, with every stateful processor
replayed from its initial state along $h$. Sequential decoding generates a trace $T=(t_1,\dots,t_n,\omega)$; the
engine returns $\rho_E(T)$. The speculation-off law of returned outputs is $P=\rho_E\#\Law_K(T)$, and $\mu$ is the
law of returned outputs with speculation on. Four premises make speculative sampling exact for model-free
drafters: (i)~acceptance uses the proposal law actually used; (ii)~acceptance, residual and bonus steps are
implemented correctly; (iii)~the verification randomness of each block position is fresh: it is independent
across positions and of the randomness of logically earlier positions, and nothing decided before the position
(its target, its draft, whether it is reached) depends on it, although the engine may draw it early; and (iv)~the
draft is a deterministic function of the committed prefix, so the proposal is a point mass. Processors are
deterministic functions of their inputs and state, and the stop rule is a deterministic function of the committed
prefix. The \emph{logical prefix} of a sampled position is the prompt, the tokens committed before the current block
began, and the block's drafts before that position.

\begin{lemma-plain}[conditional composition]\label{app:composition-restated}
Assume (i)--(iv) and the following. Every reached verification position, including the bonus position, uses the
target distribution $K_E(\cdot\mid h)$ at its logical prefix $h$; for a stateful processor, a state that is
observationally indistinguishable from the sequential state at $h$ suffices. Before every emission, including the first,
the sequential stop rule is applied to the committed prefix: the engine emits the next token if and only if the rule
continues, otherwise it stops and records the rule's termination code; tokens evaluated beyond a stop are discarded. The
returned output equals $\rho_E$ of the committed trace. Then $\mu=P$.
\end{lemma-plain}

The counterexample to a converse mentioned in the paper (Section 3, after the lemma), also used in Remark 4 below: take $L=1$ and a sequential target uniform on two
tokens; an engine that flips an internal fair coin to choose the target $\delta_0$ or $\delta_1$ violates the
first condition, yet it returns the same law.

%% file: supplement/proof-body-mlsys.tex
\section{Proof of Lemma~\ref{lem:composition}}
\label{app:proof}

\paragraph{Sequential decoding.}
Fix the engine, model, prompt and configuration. Let $V$ be the finite vocabulary and let
$h\in V^{\le L}$ denote a generated prefix; the empty prefix is $\varepsilon$. The prompt is fixed
and suppressed from the notation, so $K_E(\cdot\mid h)$ means the kernel after the prompt followed
by $h$, and $|h|$ counts generated tokens only. The kernel
$K_E(\cdot\mid h)$ is a distribution on $V$ that is a deterministic function of $h$: processors are
deterministic functions of their inputs and of their state replayed along $h$. Processors with
internal randomness or external state are outside the lemma. The stop rule
$\tau:V^{\le L}\to\{\mathsf{cont}\}\cup\Omega$ is deterministic, and $\tau(h)\neq\mathsf{cont}$
whenever $|h|=L$. Sequential decoding applies $\tau$ to the committed prefix before every draw,
including the first. It draws $t_{j}\sim K_E(\cdot\mid t_{<j})$ while
$\tau(t_{<j})=\mathsf{cont}$ and returns $T=(t_{1..n},\omega)$ with $\omega=\tau(t_{1..n})$ at the
first such stop. By the chain rule (empty products are $1$),
\begin{equation}\label{eq:seqlaw}
\begin{aligned}
&\Law_K(T)(t_{1..n},\omega)\\
&=\prod_{j=1}^{n}K_E(t_j\mid t_{<j})\,[\tau(t_{<j})=\mathsf{cont}]\\
&\quad\cdot[\tau(t_{1..n})=\omega].
\end{aligned}
\end{equation}

\paragraph{The speculative engine.}
Block $b$ starts at the committed prefix $h_b$ and obtains the drafts
$d^{(b)}=\mathrm{Draft}(h_b)\in V^{k_b}$, $k_b\ge 0$, deterministically (premise~(iv)). Its
\emph{positions} are $(b,i)$ for $i=1,\dots,k_b+1$, where $i=k_b+1$ is the bonus position. Since
drafts are deterministic, $k_\ast=\max\bigl(\{0\}\cup\{|\mathrm{Draft}(h)|:h\in V^{<L}\}\bigr)$ is finite (and $\Sigma=\varnothing$ when $L=0$), and all
random variables live on the fixed, finite set of \emph{potential} positions
$\Sigma=\{(b,i):1\le b\le L,\ 1\le i\le k_\ast+1\}$, ordered lexicographically. Every
$\sigma\in\Sigma$ carries a verification variable $\xi_\sigma$ whether or not it is used. Position
$\sigma$ has target $p_\sigma$ and draft $d_\sigma=d^{(b)}_i$; at a bonus position the draft is a
sentinel $\bot$ and the target is the engine's actual bonus target; at a potential position that
does not exist for the realized block, the draft is $\bot$ and the target a fixed dummy
distribution (such a position is never reached). Its
output is $\mathrm{out}_\sigma=g(p_\sigma,d_\sigma,\xi_\sigma)$, where $g$ is the engine's step
rule. Premises (i)--(iii) are the following two requirements.
\begin{itemize}
\item[(R1)] \emph{Correct step.} For every fixed distribution $p$ and draft $d$,
$g(p,d,\xi)$ has law $p$. Under sample-and-match, $g$ draws $y\sim p$ from $\xi$ and accepts if
$y=d$. Under accept/residual with a point mass on $d$, $g$ accepts with probability $p(d)$. On
rejection it draws from the residual $\max(p-\delta_d,0)$, normalized to $p(\cdot)/(1-p(d))$ on
$V\setminus\{d\}$. So $d$ is emitted with probability $p(d)$ and every $y\neq d$ with probability
$(1-p(d))\,p(y)/(1-p(d))=p(y)$; if $p(d)=1$ the draft is accepted surely. The bonus position draws
$y\sim p$. In both rules a position accepts exactly when it emits its draft.
\item[(R2)] \emph{Fresh randomness.} The variables $(\xi_\sigma)_\sigma$ are mutually independent.
Whether $\sigma$ is reached, its draft, and every token committed before $\sigma$ are
functions of $(\xi_{\sigma'})_{\sigma'<\sigma}$ alone, and so is the target on the event that $\sigma$ is
reached. Targets computed for positions that are never reached play no role and are unrestricted. This is premise~(iii): the
engine may draw $\xi_\sigma$ early, but nothing decided before $\sigma$ may depend on it. The
conditioning is on the randomness of logically earlier positions, not on everything physically
computed.
\end{itemize}
Block $1$ starts at $h_1=\varepsilon$. A position is \emph{reached} only while generation
continues, that is, only if $\tau$ of the current committed prefix is $\mathsf{cont}$; a reached
position emits its output. Position $(b,1)$ is reached when block $b$ starts and generation
continues. Position $(b,i+1)$ is reached when $(b,i)$ was reached, $i\le k_b$, $(b,i)$ accepted,
and generation still continues. When a reached position rejects, or is the bonus position, block
$b+1$ starts at the new committed prefix. The committed tokens $C_1,\dots,C_\nu$ are the outputs of
the reached positions $\sigma_1<\dots<\sigma_\nu$ in order, where
$\nu=\inf\{n\ge 0:\tau(C_{1..n})\neq\mathsf{cont}\}\le L$. The rule $\tau$ is evaluated only on
these live prefixes, and $\sigma_{n+1}$ is defined only on $\{\nu>n\}$. For the proof we set
$C_j=\dagger$ for $j>\nu$. The
hypotheses of the lemma are:
\begin{itemize}
\item[(a)] \emph{Target.} If $\sigma_{n+1}=(b,i)$, then $p_{\sigma_{n+1}}=K_E(\cdot\mid h)$ for its
logical prefix $h=h_b\,d^{(b)}_{<i}$. For a stateful processor, a state that yields the same
processed distribution as the sequential state at $h$ for every model output suffices.
\item[(b)] \emph{Termination.} Before every emission, including the first, the engine evaluates
$\tau$ on the committed prefix. It emits $C_{n+1}$ if and only if $\tau(C_{1..n})=\mathsf{cont}$,
records $\omega=\tau(C_{1..\nu})$, and discards every token evaluated after $C_\nu$.
\item[(c)] \emph{Return.} The engine returns $\rho_E(C_{1..\nu},\omega)$.
\end{itemize}

\paragraph{Logical prefix.} If $\sigma_{n+1}=(b,i)$, then positions $(b,1),\dots,(b,i-1)$ were
reached and accepted, so they emitted $d^{(b)}_{<i}$. Block $b$ started at $h_b$, so
$h_b\,d^{(b)}_{<i}=C_{1..n}$. By (a), $p_{\sigma_{n+1}}=K_E(\cdot\mid C_{1..n})$.

\paragraph{Proof.} Fix $n\ge 0$, $h\in V^{n}$ with $\Pr(C_{1..n}=h)>0$ and
$\tau(h)=\mathsf{cont}$, and $y\in V$. By (b), $C_{n+1}$ is emitted on $\{C_{1..n}=h\}$, so
\[
\begin{aligned}
&\Pr(C_{1..n}=h,\,C_{n+1}=y)\\
&=\sum_{\sigma}\Pr(E_\sigma,\ \mathrm{out}_\sigma=y),\qquad\\
&E_\sigma=\{C_{1..n}=h,\ \sigma_{n+1}=\sigma\}.
\end{aligned}
\]
Let $\mathcal{G}^{\sigma}=\sigma\bigl((\xi_{\sigma'})_{\sigma'<\sigma}\bigr)$. By (R2) and the reach
recursion, the event that $\sigma$ is reached, the number of reached positions before $\sigma$ and
their outputs are $\mathcal{G}^{\sigma}$-measurable, so $E_\sigma$ is
$\mathcal{G}^{\sigma}$-measurable; so is $d_\sigma$, and $\xi_\sigma$ is independent of
$\mathcal{G}^{\sigma}$. On $E_\sigma$, the logical-prefix argument gives $p_\sigma=K_E(\cdot\mid h)$.
Freezing the $\mathcal{G}^{\sigma}$-measurable arguments (at a bonus position the draft is
$\bot$ and the rule samples directly), (R1) gives
\[
\begin{aligned}
&\Pr(E_\sigma,\,\mathrm{out}_\sigma=y)\\
&=\mathbb{E}\bigl[\mathbf 1_{E_\sigma}\,\Pr\bigl(\\
&\qquad g(K_E(\cdot\mid h),d_\sigma,\xi_\sigma)=y\bigm|\mathcal{G}^{\sigma}\bigr)\bigr]\\
&=\Pr(E_\sigma)\,K_E(y\mid h).
\end{aligned}
\]
Summing over the finite set $\Sigma$ therefore gives
$\Pr(C_{n+1}=y\mid C_{1..n}=h)=K_E(y\mid h)$. By induction on $n$, using this identity and (b),
every trace $(t_{1..n},\omega)$ satisfies
\[
\begin{aligned}
&\Pr\bigl((C_{1..\nu},\omega)=(t_{1..n},\omega)\bigr)\\
&=\prod_{j=1}^{n}K_E(t_j\mid t_{<j})\,[\tau(t_{<j})=\mathsf{cont}]\\
&\quad\cdot[\tau(t_{1..n})=\omega].
\end{aligned}
\]
When a factor vanishes, both sides are zero, taking the first vanishing factor. This is
\eqref{eq:seqlaw}, so the committed trace has law $\Law_K(T)$. By (c),
$\mu=\rho_E\#\Law_K(T)=P$. \qed

\paragraph{Remarks.}
\begin{enumerate}
\item \emph{Block evaluation.} Engines evaluate all positions of a block in one pass and may
precompute targets for hypothetical prefixes. This is allowed, provided (R2) holds and the target
at the reached position satisfies (a).
\item \emph{Where premise~(iv) enters.} The proof uses point-mass proposals whose draft is
$\mathcal{G}^{\sigma}$-measurable. A drafter that reads uncommitted evaluated tokens, or the
position's own randomness, need not change the law. Such reads fall outside this argument only
when they violate premise~(iv) or the measurability and independence requirements of (R2). For
example, copying an evaluated token that is itself a deterministic function of the committed
prefix violates neither. Drafters that sample proposals (model-based drafters, such as the draft-model case study retained in the replication package) are covered by
the per-position theorem~\cite{leviathan2023,chen2023specsampling}, but not by this proof: their proposal randomness
would have to be indexed like $\xi_\sigma$.
\item \emph{Records.} Nothing here constrains the records $\ell_t$; defect class D4 is outside
the lemma.
\item \emph{No converse.} In the example of \S\ref{app:setting-restated}, an internal coin selects the
target $\delta_0$ or $\delta_1$ at the single position. This violates (a) and (R2), since the
target depends on randomness outside $(\xi_{\sigma'})_{\sigma'<\sigma}$. Yet the returned law
equals $P$.
\end{enumerate}

%% file: supplement/additional-details.tex
\section{Additional method and empirical details}

Section C retains the historical method and empirical details. Sections D--G describe separately frozen
revision studies: a retrospective marginal analysis, a trained-checkpoint score pilot, a returned-output event
follow-up and a processor-control matrix.

\subsection{Adapter conditions}

The oracle enumerates the sequential reference on tiny random models (vocabularies of \VocabHF{} to \VocabLCP{} tokens,
horizon \LHF{} or \LLCP{}). It teacher-forces each prefix $h$ under an adapter contract: (A1)~prompt, horizon, end-of-sequence set, stop strings and
features match the sampler's; (A2)~processors come from the engine's own factory, fresh for every prefix, and see
only the committed prefix; (A3)~stateful processors are replayed from their initial state along $h$; (A4)~the stop
map is the engine's own speculation-off stopping behavior.

\subsection{Registered mechanism simulators}

We register H-D1-phys and H-D3-postcheck (a stopping re-check sees unrolled drafts) for llama-cpp-python, and
H-D2-cfg, H-D2-synthid and H-D2-xgr for Transformers, which replay assisted decoding's exact call schedule,
including the drafter's processor calls on placeholder logits, without rollback. With every speculation-sensitive
read set to its logical value, each llama-cpp-python simulator must reproduce $P$ to $10^{-12}$ before its
configuration's draws; the Transformers replay has no such switch and was checked against the oracle on a
development configuration with a stateless chain.

\subsection{Mechanism distances and post hoc event removal}

For each of the \NumEligibleFailW{} eligible failures, the hypothesis registered for its configuration was compatible
with the samples and the null model was rejected (main-paper Table~2). The hypothesis laws' exact distances from
$P$ exceed the corresponding observed lower bounds:
$\mathrm{TV}(\mu_H,P)=\TvHLCPOneOneNine{}$ ($D=\DLCPOneOneNine{}$) for H-D1-phys on LCP-1 and \TvHLCPTwoOneNine{}
(\DLCPTwoOneNine{}) on LCP-2, \TvHHFOneSevenSix{} (\DHFOneSevenSix{}) for H-D2-cfg and \TvHHFOneSevenSeven{}
(\DHFOneSevenSeven{}) for H-D2-synthid. Post hoc, the hypothesis events matter for
detection: removing only $A_H$ and its complement from the event family, at the same per-bound level, lowers
the guidance bound from \DHFOneSevenSix{} to \DnoAHHFOneSevenSix{}, below the threshold, whereas the other
\NumEligNoAHW{} eligible failures stay detected; main-paper Section~6.3 tests events and selection under control.

\subsection{Prior reports and source persistence}

Our dated issue and pull-request searches (\QueriesA{} and \QueriesB{} queries, logged before the confirmation data)
did not find a report of either mechanism. Inspection of the recorded source snapshots found the relevant Transformers
processor classes and prompt-lookup callback unchanged in 5.20.1 and main 23292ca5; the assisted-decoding function
was unchanged from 5.18.0. In llama-cpp-python, the processor's source file was identical in 0.3.35, 0.3.36 and
main 1652066e. These checks establish source persistence at the inspected sites, not runtime behavior or repair
compatibility on the newer builds.

\subsection{Detection-scale heuristic}

At \NperArm{} draws per arm, the single-event excess detected roughly half the time ranges from \MdeMin{} to \MdeMax{} with the event's base rate; this is a post hoc heuristic at expected counts with witnesses excluded, and power over the whole event family was not evaluated.

\subsection{Small-cell yield uncertainty}

With at most five
configurations per cell, a 5-of-5 yield has a 95\% Clopper--Pearson interval of $[0.48, 1]$ (2 of 2: $[0.16, 1]$),
and a missing unit bounds the point yield by $[0.8, 1]$.

\subsection{Cause eligibility and counting}
A cause is counted only if (E1)~the feature has a documented contract that the speculation-off arm meets;
(E2)~calibration passed; (E3)~speculation was observed; (E4)~channel-S evidence is eligible; (E5)~confirmation
passed, with a source-level link to the defect; (E6)~dated issue and pull-request searches found no
report; (E7)~the behavior is not a documented limitation; (E8)~it shares no code site with another counted cause
and is counted once across versions and manifestations; and (E9)~it is assigned to an implementation family. Record-only and
crash-only findings, public issues and causes whose prior-report status is unresolved are reported but not counted.
Main-paper Table~1 separates the guarantees of these checks.

\subsection{Historical vLLM setup and source chronology}

Each family has a frozen-rule directed probe and one random configuration per campaign.
Both V0 and V1 have a plain-sampling negative control and a temperature/top-$k$/top-$p$ row-path check;
V1 adds four per-request drafting fallback checks.
Eligibility requires a speculative opportunity, labeled from verifier steps and the reference model independently
of the verdict. Channel~S omits hypothesis events ($m=20$ per campaign, 200{,}000 draws per arm);
detected directed configurations are repeated on fresh draws. With public families and no repair arm,
RQ4 measures transfer, not new defects.

The public source-derived positive families are V1 draft-row verification ignoring allowed tokens and bad
words~\cite{sw-vllm-pr19482} (fixed upstream), logit bias~\cite{sw-vllm-issue42744}
(refused with speculation since~\cite{sw-vllm-pr31982}), and minimum tokens~\cite{sw-vllm-pr32642}
(draft rows fixed upstream; our bonus-row reading is an uncounted cause related to a closed public
report~\cite{sw-vllm-pr41768}); stop strings returning token IDs beyond the stop~\cite{sw-vllm-pr47616}
(fix merged October 9, 2026); and V0 processors applied only at each sequence's first verification-scoring
row~\cite{sw-vllm-pr12537} (V0 speculation removed in 0.10.0).
Host glibc 2.26 limits native execution to 0.8.5.post1. Earlier source checks used 0.30.0 and main 0154979;
inspection on October 16 found stop-string cleanup in main ae8f41238cc7 but absent from 0.31.0.
Neither newer snapshot was re-executed.

\subsection{Source-derived processor-call schematic}

In a source-derived illustration from committed prefix $h$, prompt lookup filters two drafts on all-one
placeholder logits. Verification then calls the processor at $h$, $hd_1$ and $hd_1d_2$ before selecting the surviving
drafts. Rejecting $d_1$ cannot undo the processor updates at uncommitted prefixes; cropping the target model's
cache does not restore processor state. This schematic sequence is not a new sampled trace.

\subsection{Expanded historical comparison design}

A separate preregistration crosses two factors across \MFourCampaigns{} fresh-model/prompt campaigns:
(A) directed probes versus one randomly generated configuration per feature family, and
(B) decisions with versus without the hypothesis event $A_H$ and its complement (main-paper Section~4.3),
on the same draws.
Transformers 5.17.0 draft-model decoding carries the stateful-processor cause (guidance, SynthID);
llama-cpp-python 0.3.35 prompt lookup carries the processor-buffer cause (minimum length, prefix parity).
A cell recovers a cause when its higher-priority detected configuration confirms with the frozen repair on fresh draws.
A permutation chi-square test compares full outputs on the same draws without oracle, events or witnesses.
The primary contrast is the per-campaign difference in recovered causes between directed and random selection
with hypothesis events. Shared families, value ranges, prompts, simulators and repairs make this a comparison of
selection policies given that knowledge.

\paragraph{Workflow comparison}
A two-part preregistration selects merged Transformers fixes by a frozen rule (84 candidates, 14 included by two labelers, 6 not previously recorded by us; Part~A), then screens them with configurations written before selection.
Each eligible exposed defect receives two sessions per arm: \badc{} and differential testing
(DT: permutation test, contract predicates, source diagnosis, repair).
Each session has one LLM executor, 90 minutes and a token cap $T_{\mathrm{cap}}$ calibrated by a shakedown run
(Part~B; main-paper Section~7).

\subsection{Expanded related-work context}

\paragraph{Speculative decoding guarantees}
Speculative sampling preserves the target distribution~\cite{leviathan2023,chen2023specsampling};
Chen et al.\ qualify this as ``within hardware numerics''.
EAGLE~\cite{li2024eagle} reports preserving that distribution, and a two-sample $\chi^2$ test did not distinguish a
from-scratch temperature-only decoder from target-only sampling~\cite{chordiya2026lossless}.

\paragraph{RL rollouts}
NeMo-RL~\cite{nemorl2026}, BubbleSpec~\cite{bubblespec2026} and TLT~\cite{adaptivedraft2025} describe their acceleration
as lossless; EfficientRollout~\cite{efficientrollout2026} retains distribution-preserving techniques.
DeepSeek-V4.1 uses model-based drafts and dynamic verification lengths~\cite{deepseekv41flash2026};
DSpark shows that admission decisions must not depend on tokens not yet sampled~\cite{dspark2026}.
These works do not compare the composed implementation with its speculation-off law, and our results make no
claim about their implementations.

\paragraph{Inference-engine correctness}
A study of 929 bugs in five engines found over 35\% with non-crash symptoms and proposed a deterministic-decoding
oracle~\cite{liu2026firstlook}. Public reports cover a greedy draft-model mismatch~\cite{sw-mlc-issue2167}
and sampling parameters omitted at non-bonus verification positions~\cite{sw-vllm-issue9834}.
Ekka diagnoses silent model-stack errors against a reference~\cite{gu2026ekka}; GRIEF fuzzes vLLM/SGLang request
traces and confirms corruption by deterministic replay~\cite{zhao2026continuous}.
None of these three research studies reports its own speculative-decoding finding or sampled-output-law comparison.
Other work traces corrupted greedy batch-speculative outputs to position, mask and cache
misalignment~\cite{batchspec2025}, or finds two defects using speculation as an ``implementation null''~\cite{specnull2026}.
\badc{} tests shipping-engine output laws with contract-derived witnesses, checks supplied mechanisms and bounds
frozen repairs on fresh draws.

FlashInfer-Bench validates stochastic sampling kernels against expected masked distributions using empirical
total-variation checks and mask predicates~\cite{flashinferbench2026}. We test composed speculative engines'
prefix/state handling, supplied mechanisms and frozen repairs.

\paragraph{Sample verification}
Token-DiFR uses a trusted reference and the same random seed~\cite{difr2025}.
Its evaluation excludes speculation, though an appendix sketches verification from per-token acceptance,
resampling and bonus records. \badc{} needs neither seed replay nor those records, but uses many draws on tiny models.
Cross-engine distance estimation~\cite{crossengine2026} reports no experiment varying speculation.

\paragraph{Probabilistic software testing}
Related work covers differential testing~\cite{mckeeman1998}, probabilistic-programming systems~\cite{dutta2018probfuzz}
and flaky probabilistic/ML tests~\cite{dutta2020flex}.
Equal speculation-on/off laws form a metamorphic relation, and finite enumeration is a known exact-oracle technique.
For this domain, \badc{} supplies a kernel oracle, contract-derived witnesses and executable hypotheses made exact
by deterministic prompt-lookup drafts.

\section{Retrospective marginal-distribution analysis}
\label{supp:marginals}

This analysis was designed after the S3 outcomes and coauthor event-wise reports were known. Its complete
protocol/code/input freeze was published before calculation; that timing does not make the historical data
prospective or preregistered for these estimands. It changes no original event family or decision.

\paragraph{Data and observables}
Use the recorded S3 guidance case (HF17-6, Transformers 5.17.0) and minimum-length case (LCP-2-35,
llama-cpp-python 0.3.35), with speculation-off, faulty and existing mechanism-intervention arms.
HF has one million draws per arm; LCP has one million off/faulty and 200{,}000 intervention draws. The
original LCP off arm is retained, rather than pooling the later repair-run off draws. The 440 authenticated
count shards yield 129 categorical counts. Fixed observables have 10/7 categories for HF first token/visible
length, and 13/8/5 for LCP first token/visible length/termination. Error, empty, other, over-cap and unknown
categories remain explicit where applicable. HF supplies no independent termination label, so no such
marginal is manufactured. SynthID has no separate intervention result.

\paragraph{Complete marginal bounds}
For a categorical observable, $\mathrm{TV}(P,Q)=\max_A|P(A)-Q(A)|$, over all category unions.
Enumerate each nonempty union once modulo complement: 511, 63, 4095, 127 and 15 events, totaling $K=4811$.
Each arm/event gets one-sided Clopper--Pearson endpoints with tail $a=0.05/(6K)$, covering three arms and
both tails under one nominal simultaneous 95\% construction across all five observables/comparisons.
Event overlap does not require event-wise independence; valid binomial sampling within each arm is assumed.
For off interval $[l_0,u_0]$ and comparison interval $[l_1,u_1]$, the absolute gap is bounded by
\[
\begin{aligned}
g_L&=\max(0,l_0-u_1,l_1-u_0),\\
g_U&=\min\{1,\max(|l_0-u_1|,|u_0-l_1|)\}.
\end{aligned}
\]
Maximizing each endpoint over the union family bounds that marginal's TV. Table~\ref{tab:marginal} retains
every planned observable and both comparisons. Coverage is nominal conditional on the sampling assumptions;
original seed independence or protocol compliance is not certified. A small marginal TV does not bound the
joint or full-sequence TV. Separate coauthor event-wise allocations are not pooled with this construction.

\input{tables/tab-marginal}

The LCP intervention's first-token empirical TV (0.2844\%) exceeds its faulty/off value (0.2686\%);
unequal sample sizes and sampling uncertainty do not support calling that ordering a deterioration.
Every intervention/off lower bound includes zero. Seven deterministic fixtures passed before calculation;
a separately written validator reconstructed all 440 shards/129 counts and checked 48{,}110 rate/gap endpoints
using beta-distribution functions within absolute tolerance $10^{-10}$. It did not import the analyzer's
parsing, aggregation or interval helpers. These are executor checks, not independent engine replication.

\section{Trained-checkpoint mechanism pilot}
\label{supp:trained}

\paragraph{Fixed setup}
Use SmolLM2-135M-Instruct (134{,}515{,}008 parameters, vocabulary 49{,}152)~\cite{allal2025smollm2},
checkpoint revision \texttt{12fd25f7}\allowbreak\texttt{7366fa6b}\allowbreak\texttt{3b4b768e}\allowbreak\texttt{c3050bf6}\allowbreak\texttt{29380bac} with nine pinned input-file hashes.
Two isolated environments run Transformers 5.17.0/5.20.2 with existing Torch 2.11.0+cu128 on one RTX 3080,
FP32, eager attention, evaluation mode, deterministic algorithms and TF32 disabled. The original Torch wheel
digest is unauthenticated; installed generation/model source hashes and dependencies are pinned. Version 5.20.2
is the setup-time PyPI release, not tested GitHub main. Plain-text prompts use no chat template or vocabulary
restriction. Sampling fixes CFG 1.5, temperature 1, top-$k$ 0, top-$p$ 1, EOS/pad 2 and at most 8 new tokens;
native prompt lookup proposes at most 3 tokens with matching n-gram size 2.

Four deliberately repetitive prompts and their exact token IDs are frozen in the package, spanning a factual
sentence, numerical sequence, unusual lexical sentence and short word sequence. They are exposure cases,
not a natural-workload sample. Each version has 16 distinct seeds per prompt/arm, 128 main requests total.
Four calibration cells (prompts 0/2, off/spec) each pair an unaudited and audited request with the same seed.
Across versions this gives 256 main requests and 16 calibration calls, using 264 distinct logical-cell seeds.
The new 32-bit seed block was checked against 17{,}120 authenticated historical draw metadata files and
the reserved seed plans of the other studies, including low 32-bit aliases. No interrupted study was resumed.

\paragraph{State reconstruction and reference scores}
The wrapper occupies the native CFG instance's actual list position, calls it once and returns its unchanged
output. Pass-through candidate hooks record committed prefixes, proposals and acceptance. Candidate filtering
uses the same processor list on all-one dummy scores, advancing CFG state before target verification.
These filtering calls record exposure/state only, never target probabilities.

Reconstruct native unconditional history from the first-pass rule and subsequently appended tokens, checking
cache length, mask length/content and last cached input token at every call. At logical prefix $h$, the
intended unconditional history is the original prompt's final token followed by $h$'s generated continuation.
Uncached full-prefix forwards evaluate the same model on the actual and intended histories. Both CFG
calculations use identical incoming conditional target scores. Native-versus-actual-history TV calibrates
cache/full-prefix numerical differences; native-versus-intended-history TV measures the processor-level
counterfactual. The target's complete conditional trajectory is not recomputed.

\paragraph{Admission and witnesses}
All four calibration pairs per version must have identical returned tokens/candidate records, no request
errors, unchanged reference CPU/CUDA RNG, passing off-path history/reference checks, actual-history TV at
most $10^{-4}$, and nonzero proposals. Failure makes that version INCONCLUSIVE without main collection.
This pilot's mechanism witness is distinct from the channel-S structural witness requiring a zero-probability
contract violation. It requires a target-verification call at the current committed prefix, actual/intended
history mismatch, native/intended TV above $10^{-3}$, actual-history TV at most $10^{-4}$ and unchanged RNG.
Uncommitted-candidate or dummy-score calls cannot satisfy the primary criterion.

\input{tables/tab-trained}

All planned calls completed once with all controls passing. The first eligible main witness per version was
saved as full FP32 native/reference/actual-history/incoming-score vectors (49{,}152 categories); selection was
by frozen order, not effect size. For the Paris prompt, native history is \texttt{`` is the capital is''}
versus intended \texttt{`` is''}; actual-history numerical TV is $2.7818\!\times\!10^{-7}$ while intended-history
TV is 0.65831723756. Native cache length is 4, consistent with the recorded tokens/mask. The arrays are
byte-identical between versions for this initial checkpoint/prompt state.

The separate frozen analyzer reconstructs roster, pairings, histories/cache lengths, candidate linkage,
controls and summaries; NumPy FP64 softmax recomputes these two selected full-score TVs within $10^{-10}$.
Other effects remain logged scalars, not independently model-recomputed values. Nine deterministic fixture
tests passed per environment before collection. The complete freeze was pushed and read back before any
experimental forward, including calibration; no prompt/seed/rule changes, relaunches or missing-output
regeneration occurred. One sequential worker used 248.344 seconds including process startups, below the
frozen 1200 seconds/version and 2400 seconds total ceilings. This is execution time, not a throughput benchmark.

Repeated seeds do not add independent initial states; calls share state and are not independent trials.
The initial TVs in Table~\ref{tab:trained} are descriptive scalar extracts, with full-array rechecks only for
the first eligible witness per version. This is one small trained checkpoint, four constructed prompts,
one CFG setting and one host. This score pilot does not test final returned-text distributions, estimate
prevalence or establish production-model transfer, a new cause, end-to-end repair assurance or testing-method superiority.

\section{Trained-checkpoint returned-output event}
\label{supp:trained-event}

\paragraph{Selection and fixed setup}
This follow-up tests one event in the first token actually returned by native generation. It uses the Paris
prompt, trained SmolLM2 checkpoint and Transformers 5.20.2 environment from Section~\ref{supp:trained}, with
the same CFG and prompt-lookup settings. The prompt and event were selected using that pilot's data.
The event contains 1,021 token IDs chosen from its earlier score snapshot and fixed before this study's
fresh requests. It is a deterministic function of the returned sequence, rather than an internal processor
call or candidate score. This targeted design tests an exposed configuration; it does not estimate workload frequency.

The complete protocol, code and input freeze at commit
\texttt{760cc939}\allowbreak\texttt{d0041b45}\allowbreak\texttt{0e3742ec}\allowbreak\texttt{bcd1910c}\allowbreak\texttt{b0e88a43}
was pushed and read back before calibration or any checkpoint forward for this study. Twenty-four deterministic
fixtures passed before collection. Three fixed arms each made 1,024 requests: ordinary decoding with native CFG,
native prompt lookup with native CFG, and prompt lookup with functional prefix-history CFG. The control
reconstructs guidance from the logical prefix instead of retaining the native mutable history.
The fresh seed roster is known-disjoint from the checked historical IDs; this check does not establish
independent random streams or independence from unknown historical streams. All 3,072 main requests and
16 separate calibration calls completed once, with no error, retry, OOM or substituted request.

\paragraph{Shared nominal confidence construction}
Let $p_0$, $p_s$ and $p_h$ be the fixed returned-event probabilities for ordinary decoding, native prompt
lookup and the functional history control. Three two-sided Clopper-Pearson intervals use tail allocation
$0.05/6$ per endpoint. The union bound gives nominal 95\% simultaneous coverage of all three probabilities
when event indicators are independent and identically distributed Bernoulli draws within each fixed arm.
Cross-arm independence is not needed for that union bound. Fresh seed IDs alone do not certify the
within-arm assumptions. Both reported comparisons derive from this same probability family.

For intervals $[l_0,u_0]$, $[l_s,u_s]$, $[l_h,u_h]$, the signed gap $p_0-p_s$ lies in
$[l_0-u_s,u_0-l_s]$. The absolute control gap $|p_h-p_0|$ is at most
$\max(|l_h-u_0|,|u_h-l_0|)$. Table~\ref{tab:trained-event} gives the counts and probability intervals.
The observed ordinary-minus-native gap is 66.2109 percentage points, with simultaneous bounds
$[60.2649,71.6080]$ points. Its lower bound exceeds the frozen 5-point threshold. Since the event is a
function of returned tokens, its gap also lower-bounds the returned-sequence TV at this configuration.
It neither estimates nor upper-bounds full-sequence TV.

\input{tables/tab-trained-event}

The control's empirical absolute event gap from ordinary decoding is 0.9766 points, with simultaneous
upper bound 6.9487 points, below the frozen 10-point threshold. This is an event-specific intervention
bound. It establishes neither exact equality, a 1\% bound nor restoration of the whole returned-output law.
The functional implementation is a research control; it is not an upstream software fix. It also replaces
native caching with full-prefix recomputation. The comparison therefore does not isolate history handling
from every implementation difference or identify a unique, complete cause. It remains consistent with the
CFG state/history mechanism already counted in the historical study and adds no distinct cause.

\paragraph{Exposure and numerical controls}
Each speculative main request had a nonempty initial candidate exposure followed by a target call at the
original committed prompt. The frozen analyzer reconstructed every roster entry, token history, candidate
block and returned-token binding. Eight same-seed pairs matched returned tokens and candidate blocks when
the incremental CFG audit/reference checks were enabled. Both runs retained builder/candidate hooks, so this
checks the added wrapper on those seeds, not every instrument or main seed. Reference forwards preserved CPU/CUDA RNG. Full saved-score checks give maximum
native-versus-actual-history TV $2.7818\!\times\!10^{-7}$ and maximum ordinary-versus-functional initial-score
TV $2.3315\!\times\!10^{-6}$, both below the frozen $10^{-4}$ tolerance. Initial event-score probabilities
were constant within arms: 0.8015241528 ordinary, 0.1432049723 native prompt lookup and 0.8015222098
functional prompt lookup. These scores support interpretation; the inferential counts come from returned tokens.

One local RTX 3080 worker completed in 1,868.625 seconds, below the frozen 2,700-second ceiling, and the
descendant watchdog recorded normal completion without intervention. No interrupted study was resumed,
failed output regenerated, environment installed or model downloaded. The study reused the pinned checkpoint
and isolated environment. The original Torch wheel digest is unverified.

\paragraph{Recorded output-directory qualification}
The parent received a repository-relative output path, while the frozen child used its module directory as
the working directory. The child consequently wrote raw files under a nested repository-relative path.
This was recorded during collection, before outcome analysis. After normal completion, the 11 child output
files were copied byte-for-byte to the canonical run directory; the original tree remains present.
The lineage record binds each original/canonical path, byte count and matching SHA256. The frozen analyzer
consumed those identical bytes. This is qualified administrative relocation, rather than unqualified
execution-layout compliance. No collection code, prompt, seed, metric, sample size, control or stop rule
changed, and the frozen runner was neither patched nor rerun.

\paragraph{Evidence status and scope}
The artifact binds the request stream, calibration snapshots and source/input manifests by SHA256.
Its gzip unpacker authenticates and recovers the original recorded bytes.
A separate executor implementation recounted returned tokens and inverted binomial CDF/SF
functions; all six interval endpoints and both decisions agreed within $10^{-10}$. This is a supplementary
numerical check of delivered records, not independent engine execution. The original execution history
was not independently audited.

On the tested non-collecting WSL2 Linux host, the frozen analyzer returned \texttt{INCONCLUSIVE} on the
original published run before validating main rows: exact reconstructed-receipt equality failed on
two score-control floats, with absolute differences about $4.2\!\times\!10^{-19}$ and
$3.3\!\times\!10^{-19}$. A macOS receipt reconstruction also differed; no complete analyzer replay
was reported there. The numerical origin was not isolated. A separate authenticated recount agrees
with the original counts and endpoints but does not override the Linux replay stop. The collecting-host
result and frozen files remain unchanged. Section~\ref{supp:cross-host} distinguishes this replay
limitation from a new collection on the second host.

The follow-up adds one trained returned-output event bound. It establishes no additional model or runtime
generality, GitHub-main behavior, production relevance, testing-method superiority or RL consequence.

\input{supplement/b1d-processor-matrix}
\input{supplement/second-host-returned-event}
\input{supplement/missing-hypothesis}
\input{supplement/workflow-comparison}

%% file: tables/tab-marginal.tex
\begin{table}[t]
\centering\footnotesize
\caption{Complete categorical-marginal TVs, in percent. Parentheses give the nominal simultaneous 95\% bounds.
HF denotes guidance; LCP denotes the minimum-length probe. All five planned marginals are retained.}
\label{tab:marginal}
\begin{tabular}{@{}lrr@{}}
\toprule
Marginal & Faulty/off & Intervention/off\\
\midrule
HF first token &1.2971&0.1479\\
&$(0.8361,1.7581)$&$(0,0.5883)$\\
HF length &0.8916&0.0441\\
&$(0.4672,1.3160)$&$(0,0.4565)$\\
LCP first token &0.2686&0.2844\\
&$(0.1274,0.7162)$&$(0,1.0355)$\\
LCP length &14.1202&0.1730\\
&$(13.7314,14.5086)$&$(0,0.9210)$\\
LCP termination &13.4372&0.0907\\
&$(13.0553,13.8187)$&$(0,0.6386)$\\
\bottomrule
\end{tabular}
\end{table}

%% file: tables/tab-trained.tex
\begin{table}[t]
\centering\footnotesize
\caption{Initial committed-prefix conditional-score TVs on the trained checkpoint. The values match across
the two pinned versions. Every prompt has 16 spec requests per version, all with eligible witnesses somewhere
in the trace. These are four designed initial states, not independent repetitions over seeds.}
\label{tab:trained}
\begin{tabular}{@{}lr@{}}
\toprule
Fixed repetitive prompt &Initial conditional TV (\%)\\
\midrule
Paris factual sentence &65.8317\\
Numerical sentence &4.1911\\
Unusual lexical sentence &3.8606\\
Short word sequence &45.3831\\
\bottomrule
\end{tabular}
\end{table}

%% file: tables/tab-trained-event.tex
\begin{table}[t]
\centering
\caption{Fixed returned-first-token event at one pilot-selected trained-checkpoint prompt, with 1,024 requests
per arm. Rates and intervals are percentages. The intervals share one nominal simultaneous 95\% construction.
The three arms use native CFG with ordinary decoding, native CFG with prompt lookup, and functional
prefix-history CFG with prompt lookup.}
\label{tab:trained-event}
\small
\begin{tabular}{@{}lrrr@{}}
\toprule
Arm & Count & Rate & Interval \\
\midrule
Ordinary & 823 & 80.3711 & [77.2287, 83.2620] \\
Native lookup & 145 & 14.1602 & [11.6540, 16.9638] \\
History control & 833 & 81.3477 & [78.2583, 84.1774] \\
\bottomrule
\end{tabular}
\end{table}

%% file: supplement/b1d-processor-matrix.tex
\section{Prospective processor-control matrix}
\label{supp:processor-matrix}

This separately frozen study extends the conditional-score evidence of Section~\ref{supp:trained}
with exercised processor controls and eight fixed prompts. It changes no historical decision or cause count.
It is distinct from the one pilot-selected returned-event experiment in Section~\ref{supp:trained-event};
the new matrix does not test returned-output laws.

\paragraph{Fixed setup and prospective scope}
This matrix study's protocol (B1D), code, active roster, seeds, model identity and installed runtime were bound at commit
\texttt{b0ad6153}\allowbreak\texttt{08802ac5}\allowbreak\texttt{5e492fec}\allowbreak\texttt{d67308d5}\allowbreak\texttt{dc864861}
and published/read back before its first checkpoint forward, including calibration.
The manifest SHA256 is
\texttt{49bf2442}\allowbreak\texttt{c0252ea0}\allowbreak\texttt{57a20cc6}\allowbreak\texttt{e598b317}\allowbreak\texttt{43b57577}\allowbreak\texttt{ba8e9d27}\allowbreak\texttt{63e21f32}\allowbreak\texttt{07cffa79}.
The artifact retains the exact strings, token IDs and ordered roster in \texttt{ACTIVE\_INPUTS.json}.
Only Transformers 5.20.2 and the same pinned SmolLM2 checkpoint as Section~\ref{supp:trained} were executed;
5.17.0 and SynthID were deferred. The setup uses one local RTX 3080, FP32/eager evaluation, deterministic
algorithms, no TF32, no chat template or explicit negative prompt, and at most eight new tokens.
Temperature is 1, top-$k$ 0, top-$p$ 1 and EOS/pad 2. Native prompt lookup proposes at most three tokens
with maximum matching $n$-gram size two. The original Torch wheel digest remains unauthenticated.

P01--P04 complete repeated prose, record, code and numerical prefixes. P05--P08 request a reminder,
city extraction, a function and a numerical-pattern description from the corresponding content domains.
These pairs also change instructions and lengths, so they are not matched causal estimates of repetition.
The source-derived pre-filter lookup found initial candidates on P01--P04 and unexpected unigram
candidates on P06/P08; P05/P07 had none. All eight prompts were retained before collection.
Initial absence is not whole-request unexposure: later proposals can arise after generated tokens.

\paragraph{Roster and exercised prerequisites}
The validity unit is processor family by version, in order plain, repetition penalty 1.1, then CFG 1.5.
Plain and repetition each have ordinary/native-prompt-lookup arms; CFG has ordinary/native-prompt-lookup
arms with native state and with functional prefix history. Every prompt/arm has two main seeds.
Plain and repetition each contribute 32 main and 64 calibration calls; CFG contributes 64 main and
128 calibration calls: 128 main plus 256 calibration calls, 384 total.
The 192 logical-cell seeds occupy a separately reserved half-open block
$[2902100000,2902104096)$, checked against 359 known historical/reserved intervals.
This available-catalog overlap check does not establish independent random streams.

CFG admission requires complete valid plain/repetition units and at least one main speculative request
with nonempty native candidate exposure in each. Both controls satisfy this exercised criterion, with
14 exposed requests each. Their target scores agree with source-defined plain/repetition references
within $10^{-4}$; zero counts under the CFG-specific witness predicate alone would not establish this.
Unexposed requests remain in every planned denominator.

\paragraph{Four-level instrumentation calibration}
Each prompt/arm has one common-seed group of four separate requests: native baseline, candidate-only
hooks, processor observation, and full reference/snapshot instrumentation. Prefix-arm baseline retains
its defining functional intervention. Baseline has no audit/candidate/cache hooks; candidate-only adds
chain inspection and native candidate/cache records; observation adds prefix/history/cache observation
(an identity processor for the empty plain chain); full instrumentation adds numerical references and
predetermined snapshots. All four levels must match returned IDs and final CPU/CUDA sampling-RNG digests.
Candidate-only, observation and full levels must also match semantic candidate blocks and target-cache
events, excluding observation-specific call-range counters. All 64 groups passed before their unit's
main calls. These finite controls do not prove universal noninterference.

\paragraph{Conditional witness and intervention}
The witness requires a target-verification call at its current committed prefix, different reconstructed
actual/intended CFG histories, native/intended-history TV $>10^{-3}$, native/actual-history TV
$\leq10^{-4}$ and preserved RNG. Dummy-score candidate filtering and uncommitted prefixes are excluded.
Native history follows the actual processor-call sequence, with its cache length, input token and
attention mask checked. Intended history is the original prompt's final token followed by the generated
continuation at the current prefix. Both reference calculations reuse the same incoming conditional
scores and perform an uncached model forward on their respective unconditional histories.
The functional arm recomputes intended history with \texttt{use\_cache=False}; it changes history,
cache use and execution together, and shares its implementation with its reference calculation.
It is a bounded research control rather than a native-cache-preserving production repair.

\begin{table*}[t]
\centering
\caption{B1D main speculative matrix. Each prompt/arm has two planned and observed requests;
the aggregate denominator is 16 per arm. E counts requests with any nonempty native candidate block;
W counts requests with a qualifying CFG conditional-history witness. Plain/repetition columns report
exposure, with their numerical reference agreements assessed separately. Initial TVs are native CFG
first-target values as fractions, identical across the two seeds. Prefix CFG has 13 observed exposures;
W=0 holds by definition because actual and intended histories coincide. Its numerical consistency check
uses a shared reference implementation and can fail. Ordinary arms contribute the other 64 main requests
and have no speculative exposure; their numerical controls pass.}
\label{tab:processor-matrix}
\begin{tabular}{lrrrrr}
\toprule
Prompt & \shortstack{Plain\\E} & \shortstack{Repetition\\E} & \shortstack{Native CFG\\E/W} & \shortstack{Prefix CFG\\E/W} & \shortstack{Native CFG\\initial TV} \\
\midrule
P01 & 2 & 2 & 2/2 & 2/0 & 0.0224984 \\
P02 & 2 & 2 & 2/2 & 2/0 & 0.0347482 \\
P03 & 2 & 2 & 2/2 & 2/0 & 0.136422 \\
P04 & 2 & 2 & 2/2 & 2/0 & 0.00240589 \\
P05 & 0 & 1 & 0/0 & 1/0 & 0 \\
P06 & 2 & 2 & 2/2 & 2/0 & 0.380219 \\
P07 & 2 & 1 & 0/0 & 0/0 & 0 \\
P08 & 2 & 2 & 2/2 & 2/0 & 0.216442 \\
\midrule
All (16/arm) & 14 & 14 & 12/12 & 13/0 & --- \\
\bottomrule
\end{tabular}
\end{table*}

Native CFG witnesses occur in 12/16 planned main speculative requests, at six of eight fixed prompt
states, including all four predeclared lookup-oriented states (8/8 requests). The exposed-only count
12/12 is conditional and does not replace the planned denominator. P05/P07 remain unexposed native-CFG
nonwitnesses, not successful exercised negatives. P06/P08 retain their disclosed unigram proposals.
The 39 qualifying target calls are nested within the 12 requests, and repeated seeds add no independent
initial states. Counts and TVs are descriptive fixed-state diagnostics, not prevalence estimates.

At P01's first target, actual unconditional history is
\texttt{[5909,418,27616,5909]}, versus intended \texttt{[5909]}.
The three preceding candidate-filter calls advanced CFG before target verification.
All six witnessing states already qualify at the original-prompt first target, before later target
speculative-prefix advancement or target-cache rollback can explain the mismatch.
Native/actual-history agreement bridges the cached execution to an uncached forward on the reconstructed
history; the differing intended-history calculation supports a history-mediated conditional-score effect.
The functional intervention does not isolate every later history/cache difference.

\paragraph{Saved arrays and scalar-only coverage}
Every full-audit request saves its exact first target, plus its first strictly later eligible CFG witness
when present. Selection uses frozen order rather than maximum effect. All 206 snapshots (48 plain,
48 repetition, 110 CFG) retain finite full-vocabulary FP32 incoming, native, intended and actual-history
score arrays; CFG also retains raw actual/intended unconditional logits for guidance-formula checking.
The analyzer checks paths, SHA256, request/call identities, formula arithmetic and native/reference TVs.
A separately written local CPU-only recount checks all 384 ordered rows and all 206 saved snapshots;
saved/logged TV discrepancies are below $9\!\times\!10^{-16}$.
This checks the recorded arithmetic and snapshots, not an independent checkpoint execution.

Across calibration and main there are 1,672 full target calls with native/reference and native/actual TV
pairs: 478 plain, 466 repetition and 728 CFG. Only 206 have saved vectors; the other 1,466 pairs remain
logged scalar-only evidence. All-full scalar numerical-control maxima are 0, $9.49784\!\times\!10^{-7}$
and $1.56728\!\times\!10^{-6}$, respectively. Saved-vector native/actual maxima are 0,
$4.51481\!\times\!10^{-7}$ and $1.38720\!\times\!10^{-6}$.
The scalar maxima must not be described as full-array rechecks. References condition on the incoming
target scores; the target model's complete conditional trajectory is not independently recomputed.

\paragraph{Prior failure, retention and scope}
The preceding attempt (B1) stopped after six calibration requests and zero main requests because its
validator treated the pinned cache-crop argument as an absolute size rather than a signed removal count.
That attempt remains INCONCLUSIVE and never resumes. B1D corrected the validator/test double,
reserved fresh seeds and a new namespace, and recollected the complete roster. Its preparation followed
the failed calibration; it is not claimed to predate that observation. No historical row fills a B1D slot,
and prompts, arms, metrics and thresholds were retained. All 384 B1D calls completed once without retry,
replacement or resume, in 392.734 seconds, within the frozen unit/suite ceilings. This is elapsed
collection time, not a throughput comparison.

The result supports conditional processor-history incompatibility in this selected model/version/settings
matrix alongside exercised plain/repetition controls. It supplies no new cause, natural-workload prevalence,
returned-output-law result for these prompts, whole-law restoration, diagnostic/testing-method advantage,
production repair, broader runtime/model transfer or RL consequence.

\clearpage

%% file: supplement/second-host-returned-event.tex
\section{Second-host returned-event reproduction}
\label{supp:cross-host}

\paragraph{Shared setup and tested host}
The Section~\ref{supp:trained-event} collector was run on a second physical host with the same frozen
collector, checkpoint, Transformers 5.20.2, Paris prompt, event, CFG settings and seed roster.
The reported host used WSL2 Linux, an RTX 2080 Ti (Turing, sm\_75) and a Ryzen 9 7950X3D, with
Python 3.12.10, Torch 2.11.0+cu128, CUDA 12.8, NumPy 2.4.4 and SciPy 1.18.0.
The preregistered launcher checked the 21 frozen file entries and nine model-file digests before
collection. Retained runtime listings include post-run rechecks. Terminal excerpts and execution-history
statements are operator attestations, rather than an independent historical execution audit.

\begin{table}[t]
\centering
\caption{Second-host returned-event reproduction using the original seeds. Each arm has 1,024 requests.
First-token (First) and complete-sequence (Full) matches compare the two runs at the same seeds.
The counts reproduce the original intervals, without increasing the sampling denominator.}
\label{tab:cross-host}
\begin{tabular}{lrrr}
\toprule
Arm & Event & First & Full \\
\midrule
Ordinary & 823 & 1,024 & 1,024 \\
Native lookup & 145 & 1,024 & 998 \\
Prefix control & 833 & 1,024 & 939 \\
\bottomrule
\end{tabular}
\end{table}

\paragraph{Returned tokens and numerical agreement}
The retained stream contains 16 calibration and 3,072 main requests. The supplied frozen-analyzer output
for this new run reports all controls passed and \texttt{POSITIVE\_WITHIN\_SCOPE}.
A separate local CPU-only check authenticates the 17 delivered hashes, recovers the 3,088 raw rows,
checks the original roster and first returned tokens, and recomputes the event counts and intervals.
It also checks the eight saved calibration snapshots, which have three distinct contents.
This check does not reconstruct every per-request history/exposure or replay the complete frozen analyzer.

Table~\ref{tab:cross-host} gives the same-seed matches. All 3,072 first tokens agree. The 26 native
prompt-lookup and 85 functional-control sequence differences occur after the first token; complete
sequences contain up to eight tokens. Recomputing the bounds from these repeated counts gives the original 60.2649-point gap lower bound
and 6.9487-point control upper bound under the same within-arm sampling assumptions and shared allocation.
Seeds were reused, so the runs are not independent samples: no observations are pooled and each arm's
sampling denominator remains 1,024.

Initial event-score probabilities differ across hosts by at most $1.90\!\times\!10^{-6}$.
Cross-host TV of saved native/reference initial-score distributions is $2.28\!\times\!10^{-6}$ for ordinary
decoding, $2.62\!\times\!10^{-6}$ for native prompt lookup and $1.24\!\times\!10^{-6}$ for functional
prompt lookup. These numerical checks corroborate agreement within the stated tolerances, not bit identity.

\paragraph{Execution and inference limits}
The single reported launch used a new parent launcher and completed without retry under its preregistered
7,200-second ceiling, extended from the original 2,700 seconds. Collector and parent monotonic times were 3,793.4 and 3,800.6
seconds, while the parent's UTC timestamps span 3,721.1 seconds. The roughly 80-second discrepancy remains
unexplained; these times are not a throughput comparison. Hashes authenticate delivered bytes, not the
truth of execution-history attestations. The original-run Linux replay remains \texttt{INCONCLUSIVE}
at exact receipt equality (Section~\ref{supp:trained-event}), despite the agreeing recount.

This adds reproducibility evidence on one additional tested host for the same implementation, model and
selected event. It supplies no independent engine implementation, fresh sample pool, additional prompt/model
scope or whole-law restoration. The functional control still changes history reconstruction, cache use
and execution jointly; neither host isolates a unique cause or establishes a production repair.

\clearpage

%% file: supplement/missing-hypothesis.tex
\section{Diagnostic limits with an incomplete reference library}
\label{supp:missing-hypothesis}

\paragraph{Question and fixed simulation}
A diagnostic can only name reference classes that its library contains. We examine
this limitation using six archived, author-constructed Transformers output laws
(the no-defect null and five non-null classes) at three previously eligible tiny-model
configurations. This is offline multinomial simulation, not new engine execution.
The complete-library fold and five leave-one-non-null-class-out folds retain the
null. Each fold uses sample sizes $N=256,2{,}048,16{,}384$ per configuration and
128 evaluation panels per generating class; one panel combines three count
vectors. The fixed roster gives 11,904 calibration and 13,824 evaluation panels.
The excluded historical configuration remains excluded. Repetitions are draws
from these fixed laws, not independent defects, prompts or models.

\paragraph{Two declared decision rules}
Both rules receive the same reference laws and count panels. Their features use
only the retained members' positive-probability outcomes, plus an OTHER bin.
An adapted compatibility rule, CP-set, intersects binomial Clopper--Pearson
interval checks across those bins. Each tail uses $0.05/(2B)$, where $B$ is the total number
of bins across the three configurations. It widens intervals by a fixed 0.005 allowance
and rejects a candidate if an observed bin has zero candidate probability.
The TV-nearest rule uses mean empirical-to-reference total variation across the
three configurations. Its rejection cutoff is the maximum retained-member
123rd order statistic from 128 separate calibration panels per member. Empty
sets mean that the rule abstains from naming a class. The rules have different error and abstention
operating points; CP-set is not the full \badc{} procedure or the original mechanism-event family.

\begin{table}[t]
\centering
\caption{Missing generating classes. Each entry gives abstain/null-only/ambiguous
counts out of 128 panels. The generating class is absent, so every nonempty set
excludes truth; null-only is a missed defect. All 15 missing-class cells are shown.}
\label{tab:missing-class}
\begin{tabular}{lrcc}
\toprule
Omitted class & $N$ & CP-set & TV-nearest \\
\midrule
HF-Bplus & 256 & 128/0/0 & 0/114/14 \\
HF-Bplus & 2,048 & 128/0/0 & 128/0/0 \\
HF-Bplus & 16,384 & 128/0/0 & 128/0/0 \\
HF-Bminus & 256 & 1/48/79 & 0/124/4 \\
HF-Bminus & 2,048 & 128/0/0 & 128/0/0 \\
HF-Bminus & 16,384 & 128/0/0 & 128/0/0 \\
HF-D & 256 & 128/0/0 & 128/0/0 \\
HF-D & 2,048 & 128/0/0 & 128/0/0 \\
HF-D & 16,384 & 128/0/0 & 128/0/0 \\
HF-Kx & 256 & 128/0/0 & 128/0/0 \\
HF-Kx & 2,048 & 128/0/0 & 128/0/0 \\
HF-Kx & 16,384 & 128/0/0 & 128/0/0 \\
HF-F & 256 & 120/3/5 & 0/128/0 \\
HF-F & 2,048 & 128/0/0 & 0/128/0 \\
HF-F & 16,384 & 128/0/0 & 0/128/0 \\
\bottomrule
\end{tabular}
\end{table}

\paragraph{Outcomes and controls}
Table~\ref{tab:missing-class} shows that behavior depends on the omitted class.
For HF-F, TV-nearest returns null-only in every panel, whereas CP-set abstains in
120, 128 and 128 panels. This does not establish a generally better diagnostic:
for omitted HF-Bminus at $N=256$, CP-set abstains only once and returns 48
null-only and 79 truth-excluding ambiguous sets. The complete-library controls
in Table~\ref{tab:missing-known} show the corresponding low-sample cost. At both
larger sizes, CP-set gives 128 correct singletons in every retained-member/null
evaluation cell. The full artifact retains all 216 evaluation and 186 calibration
cells, not only the displayed examples.

\begin{table}[t]
\centering
\caption{Complete-library controls at $N=256$. Entries are correct
singleton/abstain/ambiguous-including-truth counts out of 128. No wrong singleton
or truth-excluding ambiguous set occurred in these cells. Ambiguity is not a
correct single diagnosis; for null controls it includes non-null hypotheses.}
\label{tab:missing-known}
\begin{tabular}{lcc}
\toprule
Generating class & CP-set & TV-nearest \\
\midrule
null & 46/0/82 & 121/0/7 \\
HF-Bplus & 128/0/0 & 128/0/0 \\
HF-Bminus & 0/0/128 & 128/0/0 \\
HF-D & 128/0/0 & 128/0/0 \\
HF-Kx & 128/0/0 & 122/6/0 \\
HF-F & 122/0/6 & 122/0/6 \\
\bottomrule
\end{tabular}
\end{table}

\paragraph{Interpretation and reproducibility}
A post hoc inspection of the saved counts finds that the exact-zero support veto
alone rejects all retained candidates in exactly the 120, 128 and 128 HF-F panels
where CP-set abstains. This explains that contrast without isolating an advantage
of the complete compatibility procedure; it is a record-based observation, not
a prospectively frozen component ablation. A comparator with the same support
veto or a matched operating point was not evaluated. Candidate labels identify
library distributions, not unique source-code causes.

The protocol fixed the roster and rules before the reported single simulation.
All saved counts, thresholds and decisions are retained. The archived laws and
their historical qualification were authenticated, not regenerated by new engine
runs. This study illustrates library and decision-rule limitations. It adds no
engine defect, causal localization, trained-model transfer or diagnostic
superiority claim.

%% file: supplement/workflow-comparison.tex
\section{Registered workflow comparison and its limitations}
\label{supp:workflow}

The main text reports the comparison's limited inference. This section retains the registered account,
the execution deviations and the post hoc observations. Moving this account does not change its
decisions or turn the comparison into evidence of workflow equivalence or component value.

\paragraph{Comparison with differential testing}
% Binding wording: results/m6b/ANALYSIS.md, "Claim row: b" (m6b/analyze.py), copied verbatim; then the audit's
% qualifiers (refine-logs/EXPERIMENT_AUDIT_M6B.md action items; RESULT_TO_CLAIM_M6B.md).
% E5 (.aris/p5_route_recommendation.md 6.1): the next sentence is added before the verbatim block, which is unchanged.
RQ3 compares configuration selection. A separate preregistered workflow comparison addresses added diagnostic
value (hypothesis $P_2$). Its sessions stopped at their token cap; the reported 90-minute porting time is a
censoring value. B-r5 denotes the protocol appendix recording the deviations. We retain the registered report below.
Defects were selected by a frozen rule from Transformers' merged fix history: 84 candidates; 14 included by two
labellers; 6 not previously recorded by us. The frozen battery exposed 1 of these 6; 1 of the 1 passed the
repair-validation controls, and all of them are eligible primary units ($j = 1$). The comparison covers these $n = 1$
defects, with every session that is missing or was not run bounded in both directions. Both arms used the same
executor model, 90 minutes and the same token budget per session, the same named configurations, runner and
mechanical validator. \badc{} sessions also had to port a toolkit written for Transformers 5.17.0 to each defect's
version within that budget; 0 of the 2 \badc{} sessions in the comparison ran the ported oracle and decision code
(median porting time 90 min, censored at 90). The comparison is between complete workflows, porting included, and
does not isolate any \badc{} component. Part~B's rules were finalized after the selection outcome, after the
selection draws and before any inference-based ground-truth decision. Without tools, the executor named the pull
request for 0 of the 1 defects and a changed site for 0 of the 1. \badc{} sessions produced a validated repair in 0
of 2 sessions, DT sessions in 0 of 2; no session is missing. Per-defect validated-repair counts favoured \badc{} on 0
of the 1 defects and DT on 0, and were tied on 1. The comparison is too small to tell the workflows apart. It cannot
exclude that the per-defect counts favour \badc{} on up to 97.5\% of defects in this frame, or DT on up to 97.5\%
(exact 95\% Clopper--Pearson). It is not evidence that the workflows are equivalent, or that \badc{}'s components add
or do not add value. What \badc{} adds over differential testing remains open. $P_2$ remains open. On the 1 exposed
defects we had recorded before (stratum~K), on which the \badc{} toolkit was developed, none entered the secondary
set. We draw no comparison from them. Every session stopped on crossing the token cap $T_{\mathrm{cap}}=243{,}751$ after 1.3--4.7
minutes, never reaching the 90-minute limit, and none submitted a patch or diagnosis (mechanism judges not run); neither \badc{} session ran
the ported oracle, so the 90-minute median porting time is the censoring value. Configuration order was confounded
with arm; the recall probe and judge calibration followed the sessions, against the protocol, after only the launcher's block summary (stop reasons, errors, gate) had been read, not any repair, diagnosis, detection or validation outcome (dated appendix \mbox{B-r5}).
% End of the verbatim block. Post hoc additions E6 and E6b (.aris/p5_route_recommendation.md 6.1); sources:
% results/m6b/shakedown/{pilot2,full1,full2}/sessions/*.json (BADC: agent_end, wall 7.90/10.06/9.86 min, oracle_run
% jobs 0, validated true; T_cap 10**9 at m6b/shakedown.py:197); results/m6/gt/30068/decision.json:2708,2759,2770; pilot2 max_minutes 9 (pilot2/SHAKEDOWN.json); T_cap from full2 only (results/m6b/PLAN.json:109-110,145-146)
Post hoc, three shakedown \badc{} sessions without a binding token cap (stratum-K defect; one 9-minute pilot)
produced validated repairs after 7.9--10.1 minutes without running the oracle. The compared defect changed score
records in 14{,}692 of 50{,}000 draws, with no detectable output-law change (largest event difference 0.0054,
below \TvThreshold{}; $\chi^2$ $p=0.13$). As a record-only unit it could not demonstrate channel~S,
hypothesis or confirmation value (main-paper Section~4.4).

\clearpage